\documentclass[10pt, final, journal]{IEEEtran}
\usepackage{authblk}

\usepackage{amsmath,amssymb,amsfonts}
\usepackage{textcomp}

\usepackage{authblk}

\usepackage{cite}

\usepackage{color}

\usepackage{threeparttable}
\usepackage{multirow}
\usepackage{longtable}
\usepackage{tabularx}
\usepackage{threeparttable}
\usepackage{makecell}
\usepackage[pdftex]{graphicx}
\usepackage{subfigure}
\graphicspath{{../pdf/}{../jpeg/}}
  
\usepackage{algorithm}
\usepackage{algorithmic}

\usepackage{array}

\begin{document}

\title{A Causal Intervention Scheme for Semantic Segmentation of Quasi-periodic Cardiovascular Signals}
\author{Xingyao Wang, Yuwen Li, Hongxiang Gao, Xianghong Cheng, Jianqing Li, and Chengyu Liu, ~\IEEEmembership{Senior~Member, ~IEEE}
\thanks{Manuscript received February 13, 2022. This research was funded by the National Natural Science Foundation of China (62001111, 62171123, 62071241 and 81871444), the National Key Research and Development Program of China (2019YFE0113800) and the Natural Science Foundation of Jiangsu Province (BK20200364, BK20190014 and BK20192004).
(Corresponding authors: Xianghong Cheng; Chengyu Liu.)}
\thanks{Xingyao Wang, Yuwen Li, Hongxiang Gao, Xianghong Cheng, Jianqing Li and Chengyu Liu are with the School of Instrument Science and Engineering, Southeast University, Nanjing, 210096, China. Xingyao Wang and Chengyu Liu are also with State Key Laboratory of Bioelectronics, Southeast University, Nanjing 210096, China (e-mails: xingyao@seu.edu.cn; liyuwen@seu.edu.cn; hx\_gao@seu.edu.cn; xhcheng@seu.edu.cn; ljq@seu.edu.cn; chengyu@seu.edu.cn).}}

\maketitle

\begin{abstract}
Precise segmentation is a vital first step to analyze semantic information of cardiac cycle and capture anomaly with cardiovascular signals.
However, in the field of deep semantic segmentation, inference is often unilaterally confounded by the individual attribute of data.
Towards cardiovascular signals, quasi-periodicity is the essential characteristic to be learned, regarded as the synthesize of the attributes of morphology ($A_m$) and rhythm ($A_r$).
Our key insight is to suppress the over-dependence on $A_m$ or $A_r$ while the generation process of deep representations.
To address this issue, we establish a structural causal model as the foundation to customize the intervention approaches on $A_m$ and $A_r$, respectively.
In this paper, we propose contrastive causal intervention (CCI) to form a novel training paradigm under a frame-level contrastive framework.
The intervention can eliminate the implicit statistical bias brought by the single attribute and lead to more objective representations.
We conduct comprehensive experiments with the controlled condition for QRS location and heart sound segmentation.
The final results indicate that our approach can evidently improve the performance by up to 0.41\% for QRS location and 2.73\% for heart sound segmentation.
The efficiency of the proposed method is generalized to multiple databases and noisy signals.
\end{abstract}

\begin{IEEEkeywords}
cardiovascular signal, semantic segmentation, QRS-complex, heart sound, representation learning, causal intervention.
\end{IEEEkeywords}

\section{Introduction}
\label{sec:introduction}
The cardiovascular signals implicate rich information about heart circulation system, including electrocardiograph (ECG), phonocardiogram (PCG) and photoplethysmographic (PPG), etc., commonly used as non-invasive means for monitoring cardiovascular system and diagnosis of organic heart disease and cardiac electrophysiological abnormalities.
For paroxysmal arrhythmia and various invisible heart diseases, long-term dynamic monitoring has become an indispensable supplement to conventional test.
Automatic analysis with these signals is crucial to alleviate workload for cardiologists, especially for long-term dynamic monitoring, such as Holter or wearable ECG.
The first and most critical step for automatic diagnosis is high-precision semantic segmentation of the physiological signal, since the error will be counted up to the subsequent stages.

In clinical applications, peculiarly in dynamic environment, the temporal physiological signals are susceptible to interference from noise and individual variability.
Due to the low dominant frequency of the target components, the state identification is always confused by intra-bandpass noise.
In the past, the researchers concentrated on the preprocessing and feature extraction of such physiological signal to improve the segmentation performance \cite{santini2019versatile, voss2018hypersampling, noman2019markov}.
The essence of these methods is to amplify the inter-state difference and discrepancy between target signals and noises, such as calculating the slope change and wavelet transforming to locate the QRS-complex and P waves in ECGs \cite{madeiro2007new, marvsanova2019advanced} and fundamental heart sounds in PCGs \cite{springer2015logistic, liu2017performance}, modeling PPGs with Gaussian functions \cite{liu2013modeling, liu2014modelling}, etc..
Nonetheless, these classic methods can only deal with static scenes with single-source noise and non-severe variations.
The recent researches indicate that the supervised machine learning methods are capable of significantly improving the segmentation performance for peudo-periodic physiological signals, and are more robust in dynamic databases \cite{wang2020temporal, cai2020qrs}.

\begin{figure}[t]
  \centering
  \begin{minipage}[b]{3.4in}
    \centering
    \includegraphics[width=3.4in]{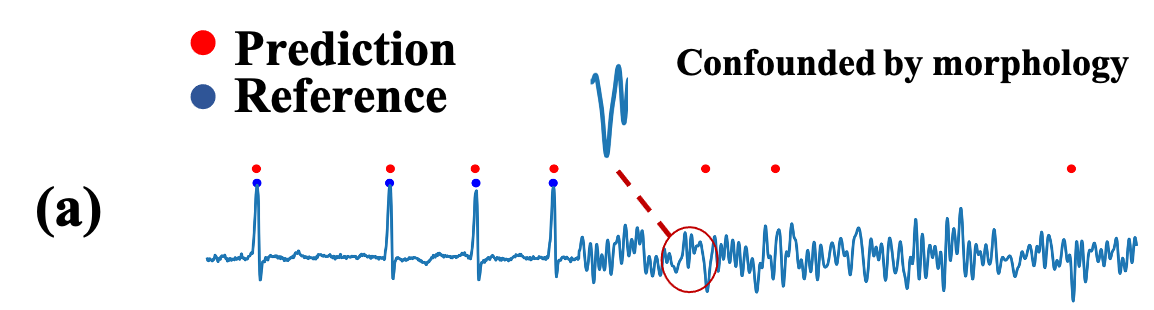} \\
    \includegraphics[width=3.4in]{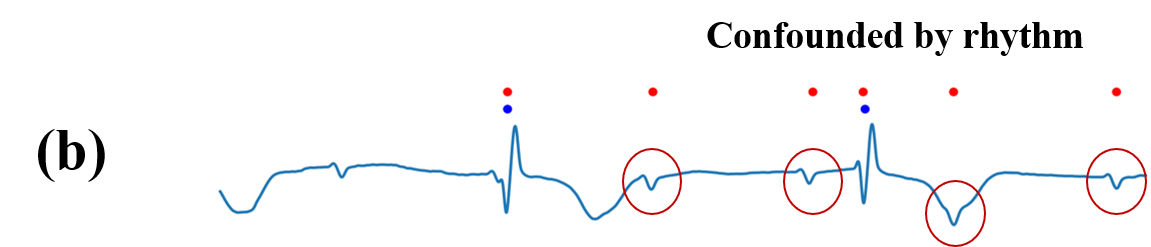}
  \end{minipage}
  \caption{Examples of QRS-complex location using deep learning method to illustrate how morphology and rhythm confound the model inference.
  (a) shows an ECG episode with noise contamination and the detection of QRS-complex is confused by the partial noisy waveforms.
  (c) is an ECG episode with III auriculo-ventricular block, which expresses unrelated rhythm of atrium (P wave) and ventricular (QRS-complex).
  Obviously, the segmented QRS-complex is misled by the atrial rhythm.}
  \label{fig:1}
\end{figure}

Pseudo-periodic is an exclusive characteristics of cardiovascular signals, which are epitomized to two attributes, attribute of rhythm ($A_r$) and attribute of morphology $A_{m}$, as follows: \\
\textbf{$\mathbf{A_{m}}$:} A signal segment of a state needs to own the general morphological characteristics of that state in all cardiovascular signals.\\
\textbf{$\mathbf{A_{r}}$:} A signal segment of a state needs to obey a repetitive pattern of that state in the same cardiovascular signal.\\

In most cases, the cardiovascular signals naturally contain these two attributes and they are mutual independent.
However, due to the abnormalities in electrophysiological activity, such as cardiac arrest, ventricular tachycardia, atrioventricular block, etc., or noise disturbances, such as leads failing, motion artefacts, etc., $A_r$ and $A_m$ would be modified.
In Fig. \ref{fig:1}, we respectively list two scenarios in QRS-complex location that $A_m$ and $A_r$ hijacks the inference of the segmentation model, respectively.

Assuming $Z$ the deep representation of an ECG episode, $A_{m}$ and $A_{r}$ should be joint dependencies of $Z$, yet are they highly coupled in the latent space, causing over-dependence of $Z$ on the onefold attribute.
In this work, we propose a solution to eliminate the individual effect from $A_m$ and $A_r$ and the intuitive thought is to intervene the attributes in latent space.


In \cite{scholkopf2021toward}, the Independent Causal Mechanisms (ICM) Principle was proposed as follows:
The causal generative process of a system's variables is composed of autonomous modules that do not inform or influence each other.
In the probabilistic case, this means that the conditional distribution of each variable given its causes (i.e., its mechanism) does not inform or influence the other mechanisms.
Applied to the segmentation of cardiovascular signals, this principle tells us that knowing one of $P(Z|A_r)$ and $P(Z|A_m)$ does not give any information about the other.

In this paper, we propose a novel contrastive learning framework combined with frame-level causal intervention for semantic segmentation of cardiovascular signals, contrastive causal intervention (CCI). 
There are four main contributions in this paper: \\
1) We establish a causal structural model to depict the implicit dependency relationship between abstracted attributes and the latent representations. \\
2) A frame-level contrastive training strategy based on the proposed CCI is designed to implement the intervention paradigm on $A_m$ and $A_r$. \\
3) We evaluate CCI on two classic tasks of cardiovascular signal segmentation, QRS location and heart sound segmentation, and comprehensive experiments for measuring the segmentation performance are implemented on a large number of independent test sets. \\
4) Additional analytical results including a real-world noise stress test and visualization of latent distributions are presented to illustrate how and why CCI improves robustness and generalization of the segmentation model.

\section{Related Work}
\label{sec:related work}
\textbf{Time Series Semantic Segmentation.}
The common Encoder-Decoder architecture for segmentation task ensures the inherent tension between semantics and location in the training process, which allows researchers to develop different variants of the Encoder structures \cite{badrinarayanan2017segnet, noh2015learning, ronneberger2015u, chen2017deeplab} for more efficient feature fusion.
According to existing researches, fully convolutional network (FCN) has been proved a superior performance in semantic segmentation task \cite{long2015fully} with controllable computational cost.
Subject to the receptive fields, the performance bottleneck has been raised due to the lack of capability for learning long-range dependency information in unconstrained scene images \cite{yang2018denseaspp} and particularly in time series \cite{qiu2021exploiting}.
To address the limited learning ability of contextual information, DeepLab and Dilation \cite{zhang2020semantic} introduce the dilated convolution to enlarge the receptive field.
Alternatively, context modeling is the focus of PSPNet \cite{zhao2017pyramid} and DeepLabV2 \cite{chen2017deeplab}.
Decomposed large kernels \cite{peng2017large} are also utilized for context capturing.
In temporal segmentation, to expand multi-scale receptive fields and leverage the inherent temporal relation, a reasonable approach is to disassemble the network into multi-branches to expand multi-scale receptive fields \cite{cai2020qrs} or distribute sub-networks at each time step \cite{hu2020temporally, jain2019accel, li2018low}.
For multi-state segmentation in pseudo-periodic signal, variants of recurrent neural network (RNN) \cite{fernando2019heart, messner2018heart} and dynamic inference \cite{jin2018improving, wang2020temporal} are utilized for learning state transition probability.

\textbf{Causal Representation Learning.}
Although methods for learning causal structure from observations exist \cite{shimizu2006linear, hoyer2008nonlinear, bauer2016arrow}, variables in a causal graph may be unobserved or unquantifiable (i.e. NN representations), which can make causal inference particularly challenging.
It is inevitable to arise statistical dependence caused by internal causal relations so that destruct performance of current machine learning methods, since the i.i.d. assumption is violated.
There has been a growing amount of efforts in performing appropriate interventions in several tasks, including image classification \cite{mao2021generative}, visual dialog \cite{wang2020visual} and scene segmentation \cite{tang2020unbiased}.
Another dilemma is the entangled factorizations in the latent space, inducing the indecomposable causal mechanisms.
Disentanglement of causal effects is crucial for introduction of structural causal models.
Recent works are concentrated on disentangled factorization in the latent space while changing background conditions \cite{locatello2020weakly, besserve2021theory}, on the basis of the invariance criterion of causal structure.

\textbf{Mutual Information Estimation}
To obtain differentiable and scalable MI estimation, recent approaches utilize deep neural networks to construct varitional MI estimators.
Barber-Agakov (BA) bound for MI \cite{agakov2004algorithm} firstly propose to approach the difficulty of computing MI by using a variational distribution.
Most of these estimators focus on MI maximization problems through providing MI lower bound.
A mainstream method is to treat MI as the Kullback-Leiber (KL) divergence between the joint and marginal distribution and convert it into the dual representation.
Based on this kernel, great efforts have been paid to explore more appropriate transformations and critics using neural networks \cite{belghazi2018mutual, nguyen2010estimating, hjelm2018learning}.
Instead of MI maximization, in this paper we explicitly use MI upper bound for MI minimization.
Most existing MI upper bounds for $I(x;y)$ require the conditional distribution $p(y|x)$ or $P(x|y)$ to be known.
Since it is unpractical in most machine learning tasks, multi variational upper bounds were explored \cite{cheng2020club}, also with a Monte Carlo approximation \cite{poole2019variational}.

\section{Method}
\label{sec:method}
\begin{figure}[]
    \begin{center}
    \subfigure[$A_m$ and $A_r$ work in a ideal causal relationship.]{
    \includegraphics[width=3.7cm, height=4.5cm]{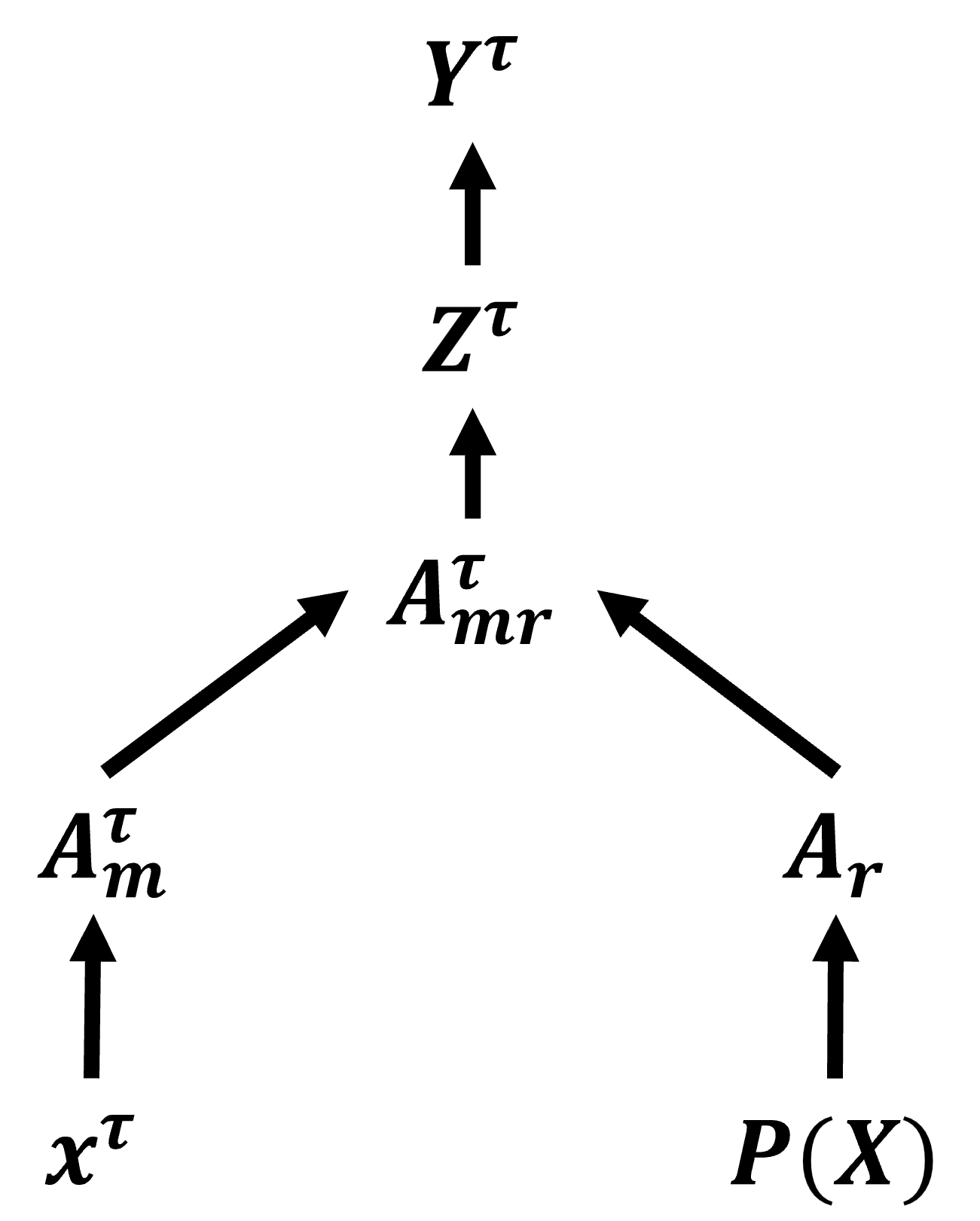}
    }
    \subfigure[$A_m$ and $A_r$ work as confounding factors.]{
    \includegraphics[width=3.7cm, height=4.5cm]{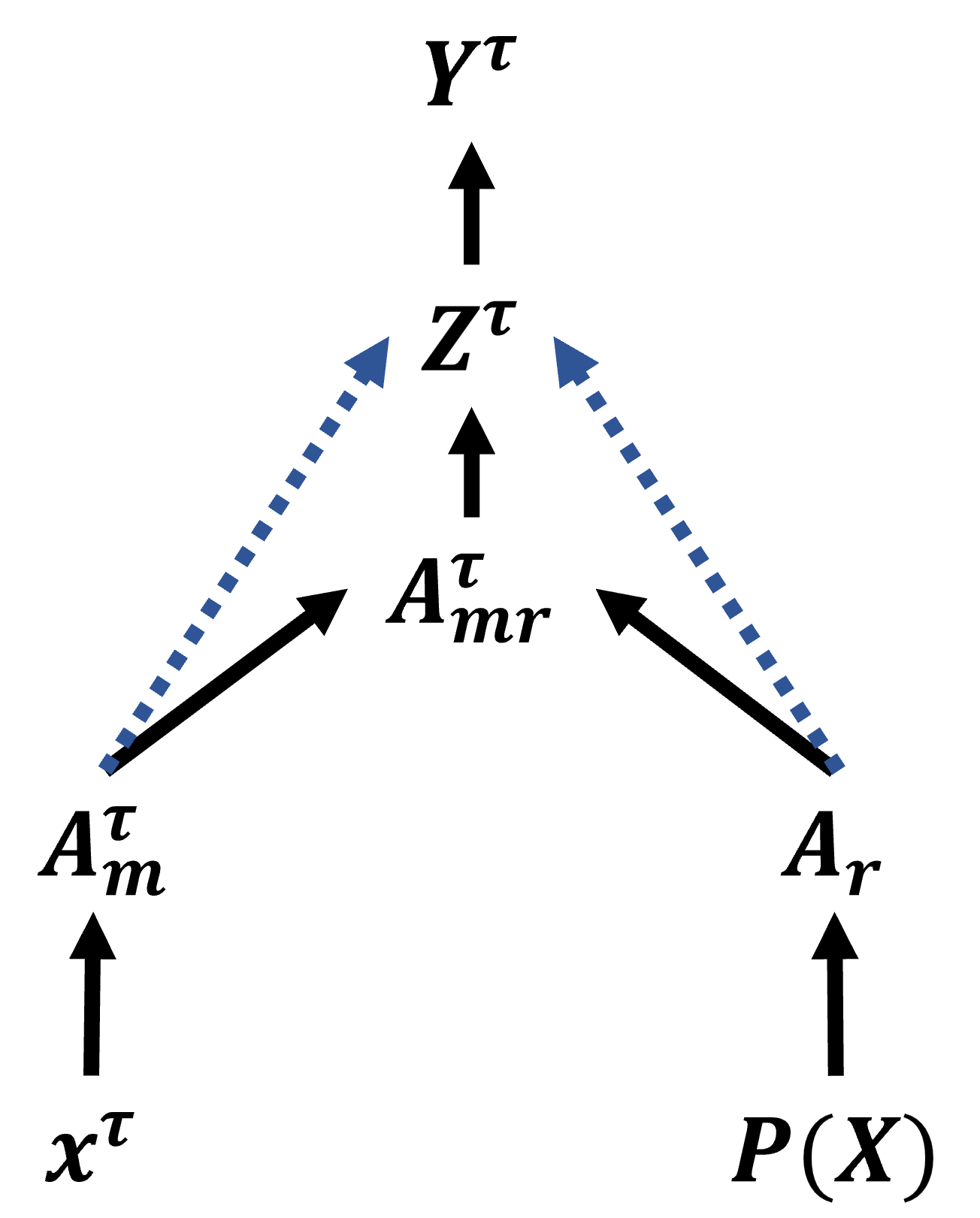}
    }
    \caption{The assumed causal structural model. (a) shows the essential causal mechanism for the attributes of morphology ($A_m$) and rhythm ($A_r$) acting on the representation $Z^\tau$ of the deep segmentation model. The dotted arrows from $A_m^\tau$ and $A_r$ to $Z_\tau$ in (b) are spurious relationships as they are not the direct dependencies.}
    \label{fig:2}
    \end{center}
\end{figure}

\subsection{Notations}

Let $X=\{x^0,x^1,...,x^\tau,...,x^T\}$ be a cardiovascular signal instance with $T$ frames and $Z=\{z^o,z^1,...,z^\tau,...,z^T\}$ be the corresponding latent feature space, where $z^\tau=\{z_{0}^{\tau},z_{1}^{\tau},...,z_{d}^{\tau}\}$ is a feature vector with $d$ dimensions.
In this paper, we focus on solving the bias problem induced by the two attributes $A_r$ and $A_m$.
A prior hypothesis is proposed as that $A_m$ is distilled from the short-term frame $x_{\tau}$ and $A_r$ from the global distribution $P(X)$.

To better understand the causal mechanism and the confounding source, we choose to insert a mediation $A_{mr}$ which defined as follows:\\
\textbf{$\mathbf{A_{mr}}$:} The morphology pattern of each type of state recurs within the same episode and obeys the data distribution of the state-specific waveform family.\\
As all the attributions are constructed in the cognitive space, we assume the ultimate $Z^\tau$ for each frame is yielded by $f(A_{rm}^\tau, U_i)$, where $U_i$ is noise, not providing any information in the latent space.
As shown in the proposed structural causal model (SCM) (Fig. \ref{fig:2}), $A_m^\tau$ and $A_r$ are the parent nodes of $A_{rm}^{\tau}$, the only dependence of the representation of the $\tau$th frame, $Z^{\tau}$.

In Sec. 3B, we intervene on $A_m$ and $A_r$ frame by frame respectively to estimate and constrain their direct effect on $z^\tau$.
Thus for each parent attribute, the do-operation will generate a signal set with $T$ variants.
Since the perturbation is adopted on each frame, we define the do-variant as $ X_{do}^{\tau}=\left\{ x^0,x^1,...,x_{do}^{\tau},...,x^T \right\} $, representing the intervention is on the $\tau$th frame. 
According to the previous definition, $A_m$ represents the local semantic information of a specific state and $A_r$ indicates the global morphology information of a recurring pattern among an episode.

\subsection{Causal Formulation}

\subsubsection{Causal Structural Model}

In Fig. \ref{fig:2}, we show the generation process of the frame-level representation $Z^\tau$ generated by the segmentation model for cardiovascular signals from the perspective of causal inference.
For most cases, $A_r$ and $A_m$ refined from biased distribution would have the probability degrade the performance of segmentation in signals that are inconsistent with the training distribution.
In the proposed SCM, we can clearly see how $A_m^\tau$ confounds $Z^\tau$ and $A_{rm}^\tau$ via the backdoor paths, $z^{\tau} \leftarrow A_m^\tau \rightarrow A_{rm}^\tau$.
Similar causal mechanism exists for $A_r$ via another backdoor path $z^{\tau} \leftarrow A_r \rightarrow A_{rm}^\tau$.
We expect to cut off the direct causal links, $A_m^\tau \rightarrow z^{\tau}$ and $A_r \rightarrow z^{\tau}$.
However, $A_r$ and $A_m^\tau$ are highly coupled in the latent space while the fitting process, and decoupling out $A_r$ and $A_m^\tau$ from representations is expensive.
Therefore, we choose to perform a constraint while fitting process to reduce the straight influence from $A_r$ and $A_m^\tau$ to $Z^\tau$, and the first step is to measure or estimate what degree the model discriminates by $A_r$ and $A_m^\tau$ in the generation process of $Z$.

\subsubsection{Causal Intervention Formulation}

\begin{figure*}[ht]
    \begin{center}
    \subfigure[An instance of intervention on $A_m$.]{
    \label{fig:3a}
    \includegraphics[width=4.5cm]{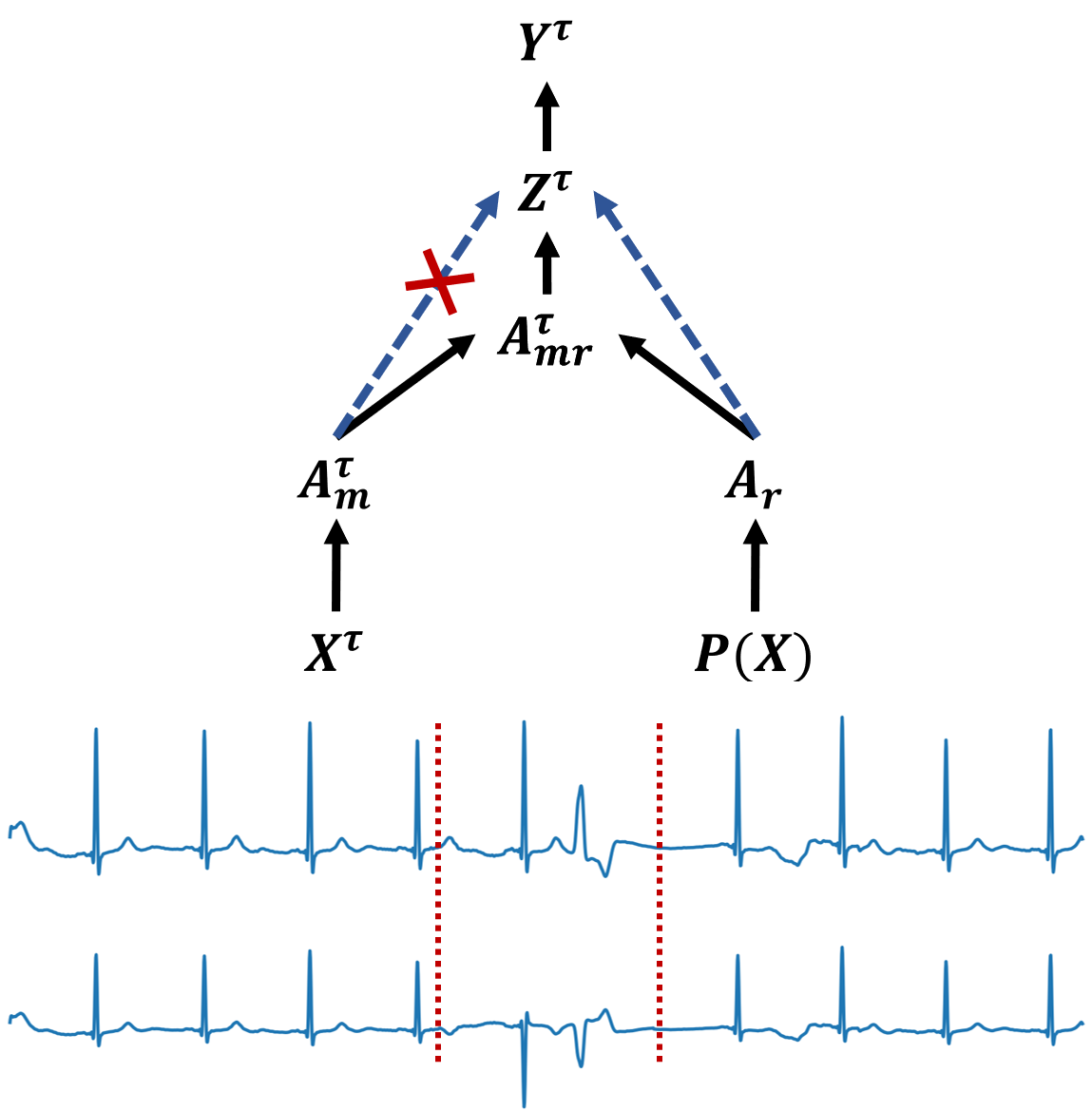}\
    }
    \subfigure[An instance of intervention on $A_r$.]{
    \label{fig:3b}
    \includegraphics[width=4.5cm]{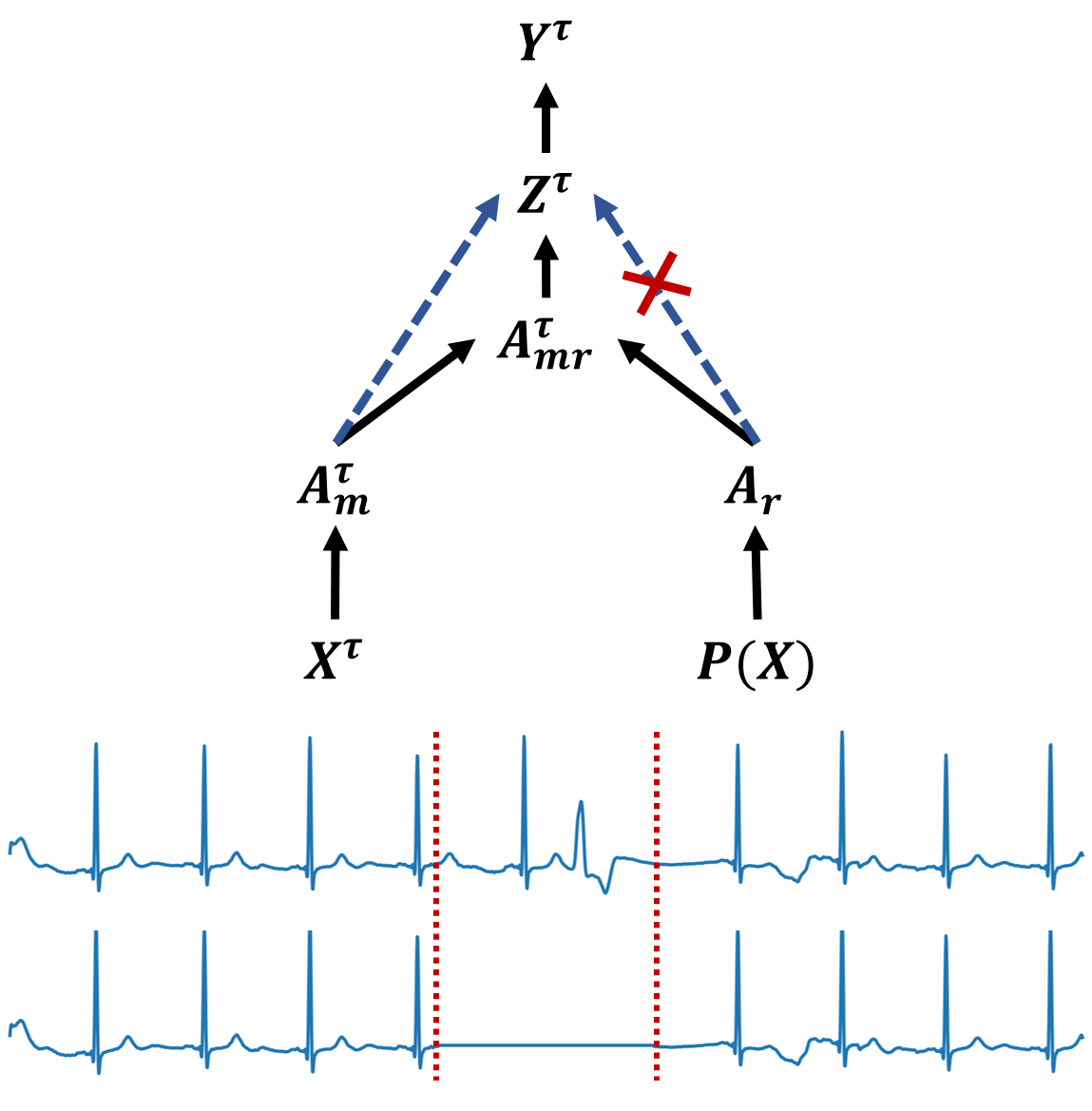}
    }
    \subfigure[Framework for contrastive causal intervention (CCI).]{
    \label{fig:3c}
    \includegraphics[width=6.5cm]{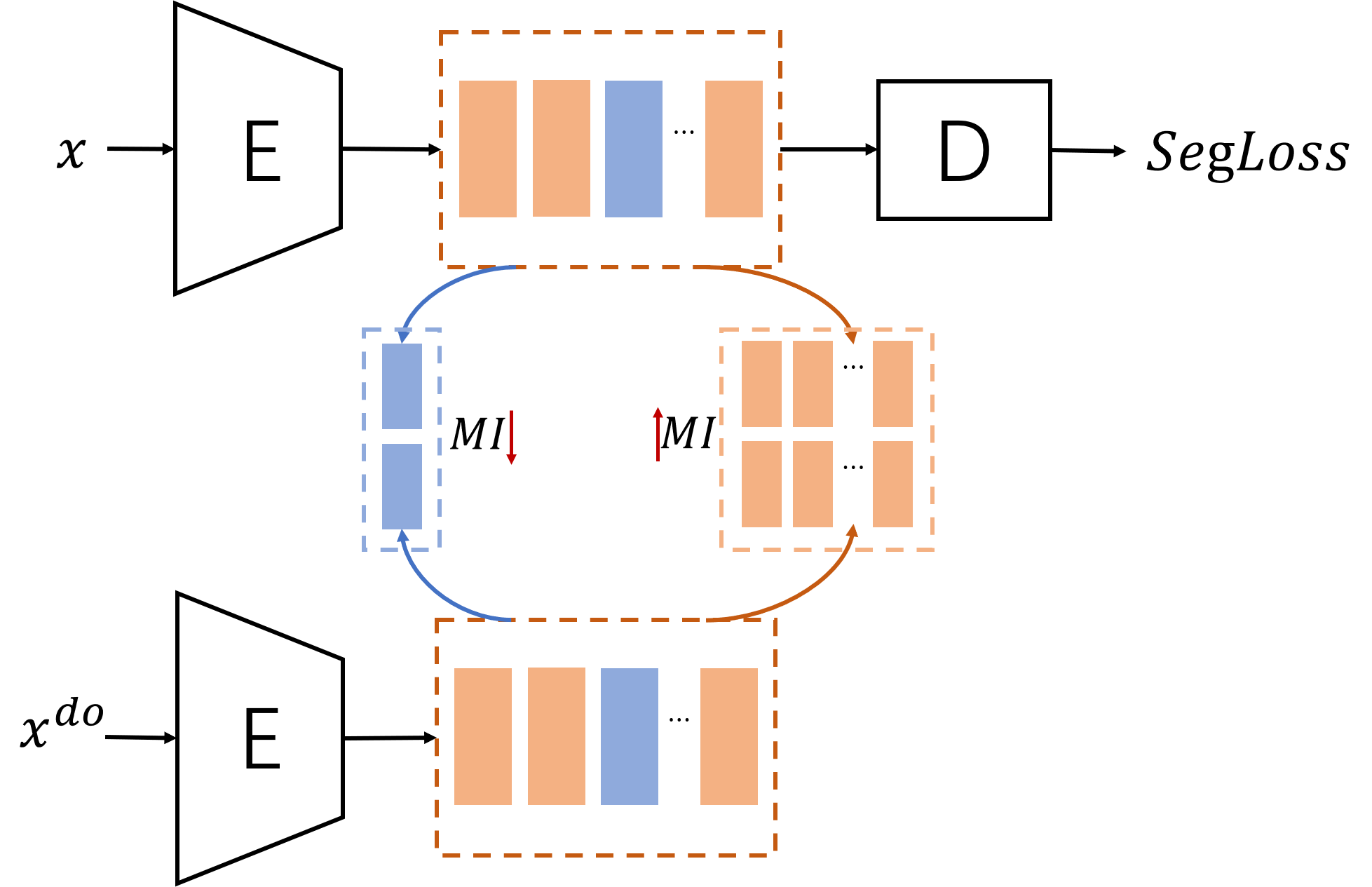}
    }
    \caption{Illustration of the intervention schemes for two attributes and the implementation method based on a frame-level contrastive framework.}
    \label{fig:3}
    \end{center}
\end{figure*}

It is unsteady to do condition on $A_{mr}^\tau$ as the $A_{mr}^\tau \rightarrow Z^{\tau}$ is confounded by $A_r$ and $A_m^\tau$, thus a more reasonable manner is to intervene on it.
We show a case study for estimating the direct effect of $A_m^\tau \rightarrow Z^{\tau}$ in the following content (the same goes for $A_r \rightarrow Z^{\tau}$ relation).
According to the ICM principle, there are two sides of conceptions when observe whether the causal link $A_m^\tau \rightarrow Z^{\tau}$ exist, which are shown as follows:\\
\textbf{a.} The statistical distribution of $z^{\tau}$ should not be varying with the change of $A_{m}^\tau$ while holding $A_{rm}^\tau$ steady.\\
\textbf{b.} The statistical distributions of $z^{\tau}$ with mutual independent $A_{mr}^\tau$ should be irrelevant even with a steady $A_{m}^\tau$.

For \textbf{a}, we can fabric the conditional distribution of $Z^{\tau}$ through changing $A_{m}^\tau$ from $a_m^\tau$ to $a_m^{\tau\prime}$, which is defined as:
\begin{align}
    \label{eqn:1} &P_{\theta}^{do(A_m)} = P_{\theta}\left( Z_{\tau} \mid do( A_{m}^{\tau} = a_m ), do( A_{mr}^{\tau}=a_{mr} ) \right), \\
    \label{eqn:2} &P_{\theta}^{do(A_m^\prime)}=P_{\theta}\left( Z_{\tau} \mid do( A_{m}^{\tau} = a_m^\prime ), do( A_{mr}^{\tau}=a_{mr} ) \right).
\end{align}

Since there is no backdoor path from $A_{m}^\tau$ to $z^{\tau}$ and $A_r$ is another confounder for $A_{mr}^\tau \rightarrow z^{\tau}$, we can block the other backdoor path through adjusting $A_r$, which gives:
\begin{equation}
    \label{eqn:3} 
    \begin{split}
        &P_{\theta}^{do(A_m)} = \\
        &\int_{a_{r}} P_{\theta}\left( Z_{\tau} \mid A_{m}^{\tau} = a_m, A_{mr}^{\tau}=a_{mr}, A_r = a_r \right) P\left( A_r=a_r \right), \\
    \end{split}
\end{equation}
\begin{equation}  
    \label{eqn:4} 
    \begin{split}
        &P_{\theta}^{do(A_m^\prime)} = \\ 
        &\int_{a_{r}} P_{\theta}\left( Z_{\tau} \mid A_{m}^{\tau} = a_m^\prime, A_{mr}^{\tau}=a_{mr}, A_r = a_r \right) P\left( A_r=a_r \right).
    \end{split}
\end{equation}

According to \textbf{a}, $P_{\theta}^{do(A_m)}$ and $P_{\theta}^{do(A_m^\prime)}$ should be consistent, inducing the objective function:
\begin{equation}
    \label{eqn:5} 
    L_m = \min_{\theta} D_{KL}\left( P_{\theta}^{do(A_m)}, P_{\theta}^{do(A_m^\prime)} \right).
\end{equation}

Unfortunately, $A_r$ is an abstract attribute with an infinite distribution and there is no feasible way to traverse the whole $A_r$ space.
Thus adjusting $A_r$ to block $z^{\tau} \leftarrow A_r \rightarrow A_{mr}^\tau$ is unprocurable.
Conception \textbf{b} provides an inverse logic to hold $A_{m}^\tau$ steady instead of $A_{mr}^\tau$, that is the statistical characteristics of $Z^\tau$ depend only on $A_{mr}^\tau$ regardless of whether $A_m$ changes.
Intuitively speaking, $A_{m}^\tau$ provides no direct information for $Z^\tau$.

If we choose to intervene on $A_{m}^\tau$, since there is no backdoor path from $A_{m}^\tau$ to $Z^\tau$ in the model, hence we can replace $do(a_{m})$ with simply conditioning on $a_m$.
The conditional distributions of $Z_{\tau}$ are given as follows:
\begin{equation}
    \label{eqn:6} 
    \begin{split}
        P_{\theta}^{do(A_m)} &= P_{\theta}\left( Z_{\tau} \mid do( A_{m}^{\tau} = a_m ), A_{mr}^{\tau}=a_{mr} \right) \\
        &=P_{\theta}\left( Z_{\tau} \mid A_{m}^{\tau} = a_m, A_{mr}^{\tau}=a_{mr}, A_r=a_r \right),
    \end{split}
\end{equation}
\begin{equation}
    \label{eqn:7} 
    \begin{split}
        P_{\theta}^{do(A_m^\prime)} &= P_{\theta}\left( Z_{\tau} \mid do( A_{m}^{\tau} = a_m ), A_{mr}^{\tau}=a_{mr}^{\prime} \right)\\
        &=P_{\theta}\left( Z_{\tau} \mid A_{m}^{\tau} = a_m, A_{mr}^{\tau}=a_{mr}^{\prime}, A_r=a_r^{\prime} \right).
    \end{split}
\end{equation}

We expect the representation of the target frame with different $A_{mr}^\tau$ should not derive correlation induced by the invariant $A_m^\tau$.
Here MI is adopted to measure the degree of correlation of the two representations and minimized as a constraint on training.
The object function is defined as:
\begin{equation}
    \label{eqn:8}
    L_m = \min_{\theta} I(P_{\theta}^{do(A_m)}, P_{\theta}^{do(A_m^\prime)}).
\end{equation}

Symmetrically, we can draw the paradigms for intervention on $A_r$ and the corresponding object function as:
\begin{align}
    \label{eqn:9} P_{\theta}^{do(A_r)} &= P_{\theta}\left( Z_{\tau} \mid A_r=a_r , A_{mr}^{\tau}=a_{mr}, A_{m}^{\tau} = a_m \right), \\
    \label{eqn:10} P_{\theta}^{do(A_r^\prime)} &= P_{\theta}\left( Z_{\tau} \mid A_r=a_r, A_{mr}^{\tau}=a_{mr}^{\prime}, A_{m}^{\tau} = a_m^{\prime} \right), \\
    \label{eqn:11} L_r &= \min_{\theta} I(P_{\theta}^{do(A_r)}, P_{\theta}^{do(A_r^\prime)}).
\end{align}

\subsubsection{Intervention Scheme}

In this section, we take scenarios of QRS location in an ECG episode to illustrate how to intervene on the two attributes, $A_m$ and $A_r$.
The first step is to define the associated physical transformation with the controlled do-operations.
As previously mentioned, $A_m^\tau$ indicates the state distribution of the local waveform morphology and $A_r$ the global recurring pattern distribution.

For do-operation on $A_m$, according to Eqn.\ref{eqn:7}, we need to solve out how to maintain the subordinate state properties of the local morphology while changing $A_r$ and $A_{mr}^\tau$.
Here we conduct a straightforward manner of reversing phase (amplitude inversion) on the target frame (as shown in Fig.\ref{fig:3a}).
$A_r$ is a global attribute, representing how the contextual morphology pattern influence the target frame.
Reversing the QRS-complex morphology on the target frame will definitely affect $A_r$ accordingly.

For $do(A_r)$, According to Eqn.\ref{eqn:10}, we wish to alternate the state of the target frame while not changing the global recurring pattern.
Here we perform a handy intervention, that is zero setting on the chosen target frame.
As shown in Fig.\ref{fig:3b}, we simply erase the morphology information of the target frame, not introducing extraneous signals.
Since no additional morphological information is introduced, $A_r$ can be approximately regarded as invariant as $A_m$ altered. 

In practical operation, for the same cardiovascular signal, we performed the above intervention in units of a frame with fixed length. 
Assuming the signal owns $T$ frames, given binary masks $xmask$ with $T$ dimensions, $x \otimes xmask\left[ \tau \right]$ indicates that the $\tau$th frame is set to zeros.
Then we have $do^{(A_r)}\left( x \right)=\left\{ x \otimes xmask\left[\tau \right] \right\} |_{\tau=1}^{T}$ and $do^{(A_m)}\left( x \right)=\left\{ x \otimes xmask\left[\tau \right] \right\} - x \otimes ( 1 - xmask\left[ \tau \right] ) |_{\tau=1}^{T}$

\subsection{Contrastive Framework for Causal Intervention}

In the previous sections, we have confirmed to utilize MI modeling the causal interventions and the specific operations.
In this section we will establish the framework so that the intervention of the target frame can form effective constraint while training.
Here we adopt the contrastive architecture with a shared-weights Encoder (E) and a Decoder (D), where we should learn representations from E to separate (contrast) original samples and intervened samples.
The designed temporal contrastive learning module is shown in Fig.\ref{fig:3c}.

The general contrastive loss is designed to learn feature representation for positive pairs to be similar, while pushing features from the randomly sampled negative pairs apart.
Unlike the conventional contrastive paradigm, the proposed contrastive method should weaken the relevance between representations before and after the intervention, namely the negative pairs in the classic contrastive conception.
According to the assumed attributes and causal inference, the frames beside the intervened target frame share the same attribute $A_{mr}$ and should own the consistent distributions in the latent space.
Thus we deemed these pairs of untreated frames as the positive pairs.
The ultimate contrastive paradigm should be:
\begin{equation}
\label{eqn:12}
    \min _{E, D} \mathcal{L}_{Seg} + \lambda_{1} \mathrm{I}\left(\boldsymbol{z}_{\tau}, \boldsymbol{z}_{\tau}^{do}\right) - \lambda_{2} \frac{1}{T-1} \sum_{i=0 , i\neq \tau}^{T} \mathrm{I}\left( z_i, z_i^{do} \right)
\end{equation}

Suppose the optimal representations of the frames with the same $A_{mr}^\tau$ should be completely consistent, i.e., $P(Z_i=Z_i^{do})=1$, then maximizing the MI between these frames can be substituted by cosine similarity distance:
\begin{equation}
\label{eqn:13}
    \max_{E} \mathrm{I}\left( z_i, z_i^{do} \right) \iff \max_{E} \frac{z_{i}}{\left\|z_{i}\right\|_{2}} \cdot \frac{z_{i}^{do}}{\left\|z_{i}^{do}\right\|_{2}}
\end{equation}

\subsection{Mutual Information Upper Bound Estimation}

Denote $P_{E}(Z|X)$ the distribution of the encoded representation for the original signal, and $P_{E}(Z|X^{do})$ the representation for the intervened signal.
For convenience, we apply $P(Z)$ and $P(Z^{do})$ representing $P_{E}(Z|X)$ and $P_{E}(Z|X^{do})$, respectively.
The proposed approach to estimate MI upper bound follows Contrastive Log-ratio Upper Bound (CLUB) \cite{cheng2020club}, which estimates MI through narrowing the gap of conditional probabilities between positive and negative sample pairs.

Difference exists that the intervention is operated frame by frame.
For the whole do-operations of the same attribute of a signal episode, the conditional distributions $P(Z \mid Z^{do})$ should be uniformed since they are homogeneous.
According to CLUB, a certified unbiased MI upper bound estimation is proposed with $N$ sample pairs ${(z_i, z_i^{do})}_{i=1}^N$ and $T$ frames for each $z_i$ as follows:
\begin{equation}
    \label{eqn:14}
    I_{CCI} = \frac{1}{N}\frac{1}{T} \sum_{i=1}^{N}\sum_{\tau=1}^{T} \left[ \log p\left( z_{i\tau} \mid z_{i\tau}^{do} \right) - \log p\left( z_{k_{i}^{\prime}\tau} \mid z_{i\tau}^{do} \right) \right].
\end{equation}

Unfortunately, $p(z_i \mid z_i^{do})$ is unknown so that a variational approximation of the distribution is given as $q_\theta(z_i \mid z_i^{do})$.
Thus, we have the variational upper bound estimation for $I_{CCI}$:
\begin{equation}
    \label{eqn:15}
    I_{vCCI} = \frac{1}{N}\frac{1}{T} \sum_{i=1}^{N}\sum_{\tau=1}^{T} \left[ \log q_\theta\left( z_{i\tau} \mid z_{i\tau}^{do} \right) - \log q_\theta\left( z_{k_{i}^{\prime}\tau} \mid z_{i\tau}^{do} \right) \right].
\end{equation}

The prerequisite for the establishment of $I(Z ; Z^{do}) \leq \mathrm{I}_{vCCI}$ is proved to be:
\begin{equation}
    \label{eqn:16}
    K L\left(p(z_i, z_i^{do}) \| q_{\theta}(z_i, z_i^{do})\right) \leq K L\left(p(z_i) p(z_i^{do}) \| q_{\theta}(z_i, z_i^{do})\right), 
\end{equation}
where $q_\theta(z_i, z_i^{do})=q_\theta(z_i \mid z_i^{do})p(z_i^{do})$ is the variational joint distribution induced by $q_\theta(z_i \mid z_i^{do})$.
And $K L\left(p(z_i, z_i^{do}) \| q_{\theta}(z_i, z_i^{do})\right)$ can be minimized by maximizing the log-likelihood of $q_{\theta} \left( z_i \mid z_i^{do} \right)$.
For $I_{vCCI}$, it is a cross-frame function $\mathcal{L}_{q}(\theta_{q}) := \frac{1}{N}\frac{1}{T} \sum_{i=1}^{N}\sum_{\tau=1}^{T} \log q_\theta\left( z_{i\tau} \mid z_{i\tau}^{do} \right)$.

A prior Gaussian distribution is provided to solve $q_\theta(z_i \mid z_i^{do})$.
Here we assume that $q_\theta(z \mid z^{do}) = \mathcal{N}\left( z \mid \mu(z^{do}), \sigma^{2}(z^{do})\right)$.
For Give samples ${(z_i, z_{i}^{do})}_{i=1}^N$, we denote $\mu_{i\tau}=\mu(z_{i\tau}^{do})$ and $\sigma_{i\tau}=\sigma(z_{i\tau}^{do})$. Then we have
\begin{equation}
    \label{eqn:17}
    q_{\theta}\left(z_{j} \mid z_{i}^{do}\right)= \frac{1}{T} \sum_{\tau=1}^{T} \left( 2 \pi \sigma_{i\tau} ^{2} \right) ^{-1 / 2} \exp \left \{ -\frac{\left(z_{j}-\mu_{i\tau}\right)^{2}}{2\sigma_{i\tau}^{2}}\right \}.
\end{equation}
Thus the upper bound of the MI between the origin and intervened representation can be solved while training, which is shown in Algorithm \ref{alg1} in detail.

\begin{algorithm}

    \caption{Training Procedure for CCI}
    \label{alg1}
    \begin{algorithmic}[1]
        \REQUIRE $D$: training set
        \REQUIRE $\alpha, \beta, \lambda_{1}, \lambda_{2}$, $N$: batch size, $T$: number of frames
        \STATE Initialization: $\theta_{e}, \theta_{d}, \theta_{q}$
        \WHILE{not converge}
            \STATE Sample $\{x^i, y^i\}_{i=1}^{N}$ from $D$
            \STATE $z_i \gets f_{\theta_{e}} \left( x_i \right)$
            \STATE $\hat{y}_i \gets g_{\theta_{d}} \left( z_i \right)$
            \FOR {$\tau \gets 1$ to $T$}
                \STATE $x_{i\tau}^{do} \gets do(x_{i\tau})$
                \STATE $z_{i\tau}^{do} \gets f\left( x_{i\tau}^{do} \mid \theta_{e} \right)$
            \STATE Log-likelihood \\
            $\mathcal{L}_{q}(\theta_{q}) = \frac{1}{N}\frac{1}{T} \sum_{i=1}^{N}\sum_{\tau=1}^{T} \log q_\theta\left( z_{i\tau} \mid z_{i\tau}^{do} \right)$
            \STATE Update $\theta_{q}^{\prime} \gets \theta_{q} - \alpha \nabla_{\theta_{q}} \mathcal{L}_{q}\left( \theta_{q} \right)$ 
                \FORALL{i}
                    \STATE Sampling $k_{i}^{\prime}$ uniformly from $\left\{ 1,2,...,N \right\}$
                    \STATE $U_{i\tau} = \log q_\theta\left( z_{i\tau} \mid z_{i\tau}^{do} \right) - \log q_\theta\left( z_{k_{i}^{\prime}\tau} \mid z_{i\tau}^{do} \right)$
                \ENDFOR
            \STATE $\mathcal{I}_{ v C C I } = \frac{1}{N}\frac{1}{T}\sum_{i=1}^{N}\sum_{\tau=1}^{T} U_{i\tau}$
            \STATE $\mathcal{L}_{Seg} = \frac{1}{N} \sum_{i=1}^{N} D_{CE} \left( y_i, \hat{y}_i \right)$
            \STATE $\mathcal{L}_{Sim} = \frac{1}{N} \frac{1}{T-1} \sum_{i=1}^{N} \sum_{j=1,j \neq \tau}^{T} \mathcal{I} \left( z_{ij}, z_{ij}^{do} \right)$
            \STATE $\mathcal{L} = \mathcal{I}_{ v C C I } - \mathcal{L}_{Sim}$
            \STATE Update $\theta_{e}^{\prime} \gets \theta_{e}-\beta\nabla_{\theta_{e}} \mathcal{L} \left( \theta_{e} \right)$
            \STATE Update $\theta_{d}^{\prime} \gets \theta_{d}-\beta\nabla_{\theta_{d}} \mathcal{L}_{ S e g } \left( \theta_{d} \right)$
            \ENDFOR
        \ENDWHILE
    \end{algorithmic}
\end{algorithm}

\section{Experiments and Analysis}
\label{sec:experiments and analysis}
In this section, we conduct comprehensive experiments with the aim of answering the following three key questions. \\
Q1: What is the role of the proposed intervention approach on each attribution of cardiovascular signals (i.e., the ablation studies of our CCI)? \\
Q2: In what domain does the proposed method improve the segmentation performance (i.e., data with physiological variation and noise contamination )? \\
Q3: How does the proposed contrastive causal intervention influences the generation process of latent representations? \\

\subsection{Experiment Setups }
\subsubsection{Database}
\paragraph{QRS location} 
We use CPSC2019-Train database \cite{gao2019open} for training and five other independent QRS-location benchmarks in the experiments, including CPSC2019-Test, MIT-BIH Arrhythmia Database (MITDB) \cite{moody2001impact}, QT Database (QTDB) \cite{laguna1997database}, INCART 12-lead Arrhythmia Database (INCART) \cite{taddei1992european}.

\paragraph{Heart sound segmentation} 
We selected 100 recordings randomly from training-a in PhysioNet/CinC Challenge 2016 \cite{liu2016open} and slice them into 5-second samples for training.
The remaining recordings in training-a and other data sets including training-b\textasciitilde f and hidden test sets (Test-b\textasciitilde e, Test-g and Test-i) from Challenge 2016 are utilized for testing.
The test recordings are restructured into Set-A\textasciitilde H according to the index of the data sets.

\paragraph{Noise Stress Test}
Expect for the artificial intra-band Gaussian noise, two main noise databases are utilized in our experiments.
For QRS location, MIT-BIH Noise Stress Database \cite{moody1984noise} is chosen to test the method's robustness when facing the real-world ECG noise, including baseline wander (bw), muscle (EMG) artifact (ma), and electrode motion artifact (em).
For heart sound segmentation, we focus on the influence brought by lung sounds recorded from the electronic stethoscope simultaneously.
The lung sound samples are extracted randomly from the database constructed in \cite{fraiwan2021dataset}.

\subsubsection{Pre-procession and Post-procession}
The pre- and post- procession in the experiments is designed to be plain and unified for the backbone model training with and without CCI proposed in this paper.

\paragraph{QRS location}
Considering the energy of QRS-complex is mainly concentrated at 8-50 Hz \cite{israel2005ecg}, we perform band-pass filtering from 0.5-50 Hz as well as mean filtering on each 10-second episode.
Since the magnitudes are not uniform across databases, standardization is conducted on ECG records after filtering and the input episodes are re-sampled at 250 Hz.
The ultimate outputs of the segmentation model are activated by Sigmoid function, approximated to the probability of the corresponding time step belonging to the QRS-complex.
Thus the decision of QRS-complex is to find candidate intervals with consecutive probabilities over a fixed threshold of 0.5.
Referring to the effective refractory period in ECGs, some of the intervals will be excluded if they are less than 200 ms.

\paragraph{Heart sound segmentation}  
The majority of the frequency content in S1 and S2 sounds is below 150 Hz, usually with a peak around 50 Hz \cite{arnott1984spectral}, and murmur is around 400 Hz. 
Thus, all the heart sound recordings were downsampled into 800 Hz.
Moreover, different digital stethoscopes vary widely in response of heart sound and noise.
Therefore, we adopted an adaptive local Wiener filter proposed in \cite{wang2020temporal} to suppress the in-band noise from system and increase the amplitude resolution of alternate segments between heart sound states.
The outputs of the models are functioned by the Softmax activation and the the time step is assigned the state with the maximal probability.
We only determine the onsets of S1, systole, S2 and diastole by positioning the alternating time steps.

\subsubsection{Evaluation Metrics}

For evaluation, Sensitivity ($Se$), positive predictive rate ($P_+$), error rate ($Er$) and $F_1$ are calculate in all databases.
These metrics are defined as follows:
\begin{eqnarray}
  Se = \frac{TP}{TP+FN} \times{100\%},\\
  P_+ = \frac{TP}{TP+FP} \times{100\%},\\
  Er = \frac{FP+FN}{TP+FP+FN} \times{100\%},\\
  F_1  = \frac{2\times{SE}\times{P_+}}{SE+P_+} \times{100\%},
\end{eqnarray}
where $TP$ is true positive, $FP$ is false positive and $FN$ is false negative. The standard grace period of 150 ms is used for beat-by-beat comparison in QRS location \cite{ec571998testing} and 100 ms for state-by-state comparison in heart sound segmentation \cite{springer2015logistic}.

\subsubsection{Implementation Details}
In this work, the Decoders for the two tasks are fixed with two-layer dense block.
For Encoder, we compare various baseline networks with 1D convolution, including DenseNet \cite{iandola2014densenet}, TCN \cite{bai2018empirical} and SE-TCN \cite{hu2018squeeze}.
Meanwhile, a multi-branch 1D convolutional neural network (MBCNN) architecture is adopted as the backbone method to comprehensively analyze CCI's performance.
MBCNN can distribute varying receptive fields into different braches as necessary to merge the full contextual information and avoid bloating due to long sequence inputs.

We implement all the methods on TensorFlow 2. 
The training set was sliced into 5 folds for training and evaluation.
The network and the basic training settings, including the optimizer (Adam), $\mathcal{L}_{Seg}$ (cross entropy) and the batch size, are unified for each task.
The $\lambda_1$ and $\lambda_2$ in the loss function, Eqn. \ref{eqn:12} are set to $0.1$ and $1$, respectively.
For training, an early-stopping strategy was adopted as follows: when the model failed to achieve the best validation accuracy in 20 consecutive epochs, the training is terminated.

\subsection{Q1: Ablation Studies for Causal Intervention}
To understand the assumed causal mechanisms and the respective effects of interventions on $A_r$ and $A_m$ in Eqn.\ref{eqn:5} and Eqn.\ref{eqn:11} while fitting process, we conduct ablation studies on the two pseudo-periodic segmentation tasks, QRS-complex location and heart sound segmentation.
For the multi-lead ECG records in the test sets, each lead is deemed as a single-lead ECG, sharing the ground truth of QRS locations while testing.

For the both tasks, we firstly conduct ablation studies on the attributes with the backbone Encoder, MBCNN.
Then we implement densenet, TCN and SE-TCN as different Encoders to evaluate the adaption of CCI.
The corresponding results are shown in Table \ref{tab:2}

\subsubsection{Main Results}

\begin{figure}[h]
  \centering
  \begin{minipage}[b]{3.4in}
    \centering
    \includegraphics[width=3.4in]{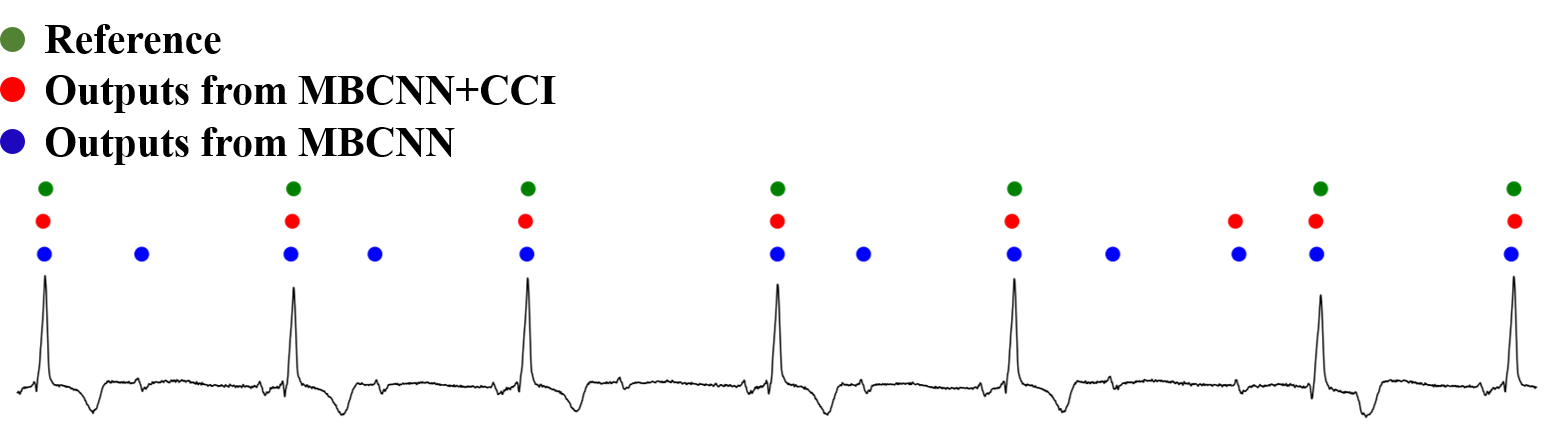} \\
    \includegraphics[width=3.4in]{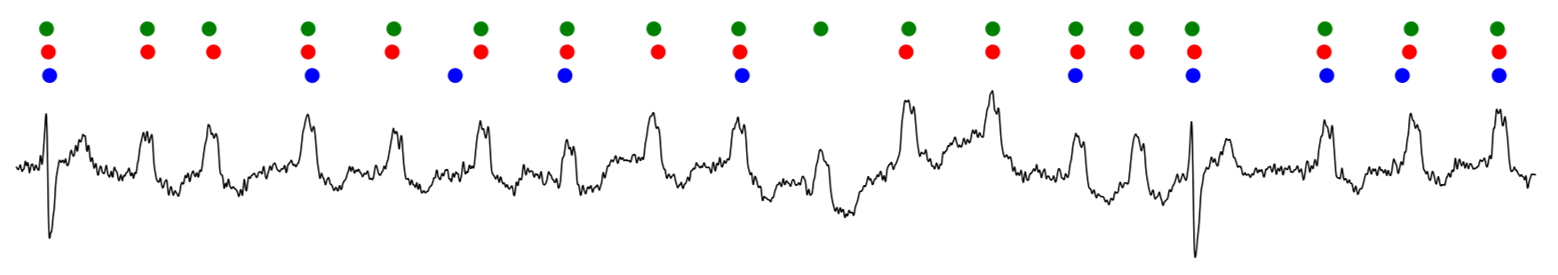} \\
    \includegraphics[width=3.4in]{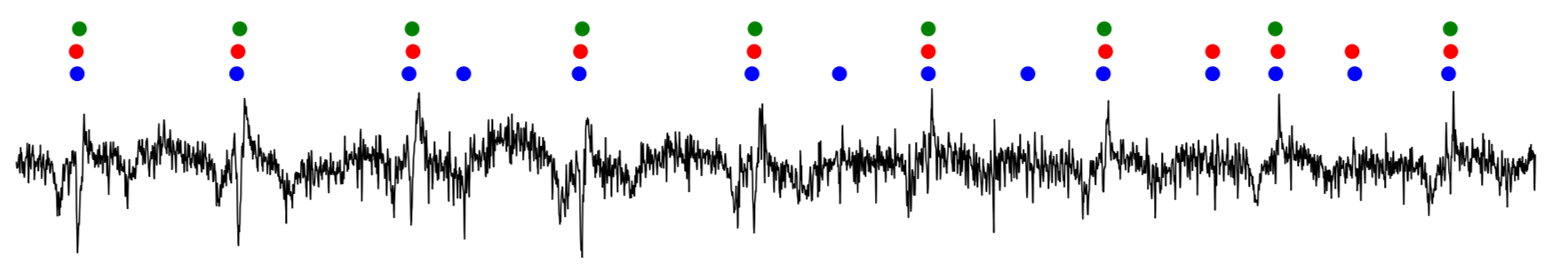} \\
    \includegraphics[width=3.4in]{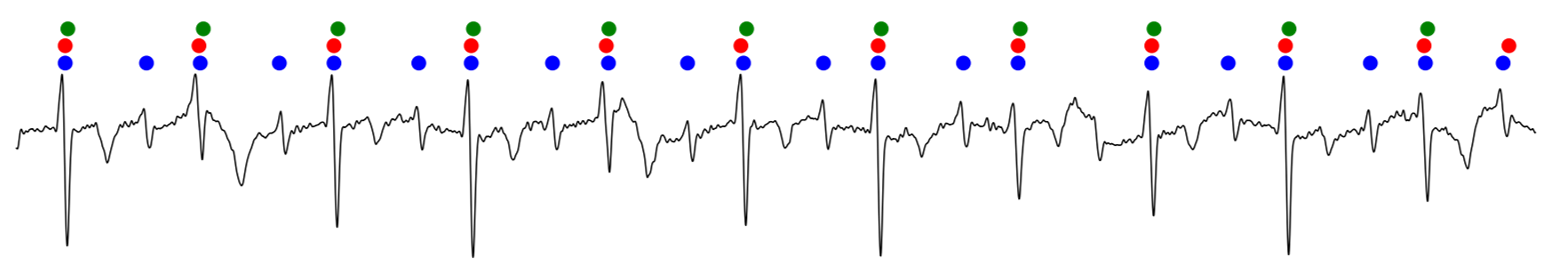}
  \end{minipage}
  \caption{The visualization results for QRS-complex location from the model trained with and without CCI. }
  \label{fig:4}
\end{figure}

\begin{table}[h]
\centering
\caption{Ablation results (\%) for QRS location with MBCNN as the backbone Encoder. The results are average of five sub models under 5-fold cross validation and the better results are \textbf{bold-faced}.
\label{tab:1}
}
\begin{tabular}{lllll}
\hline

Database      & Method & $Se$ & $P_+$ & $F_1$    \\
\hline\hline
\multirow{3}{*} {CPSC2019-Test} & MBCNN & 98.79 & 99.09 & 98.94 \\

              & MBCNN+CCI ($A_m$) & \textbf{99.32} & 99.31 & 99.31 \\
              
              & MBCNN+CCI ($A_r$) & 99.28 & 99.37 & 99.32 \\
              
              & MBCNN+CCI ($A_m$ \& $A_r$) & 99.26 & \textbf{99.45} & \textbf{99.35} \\
\cline{1-5}
\multirow{3}{*} {MITDB}  & MBCNN     & 99.20 & 99.44 & 99.32 \\

              & MBCNN+CCI ($A_m$) & 99.37 & 99.49 & 99.43 \\
              
              & MBCNN+CCI ($A_r$) & \textbf{99.41} & 99.50 & 99.45 \\
              
              & MBCNN+CCI ($A_m$ \& $A_r$) & 99.38 & \textbf{99.56} & \textbf{99.47} \\
\cline{1-5}
\multirow{3}{*} {INCART} & MBCNN     &99.35 & 99.13 & 99.24 \\
              
              & MBCNN+CCI ($A_m$) & 99.45 & 99.28 & 99.37 \\
              
              & MBCNN+CCI ($A_r$) & \textbf{99.48} & 99.23 & 99.35 \\
              
              & MBCNN+CCI ($A_m$ \& $A_r$) & 99.45 & \textbf{99.32} & \textbf{99.39} \\
\cline{1-5}
\multirow{3}{*} {QT} & MBCNN     &99.93 & 99.90 & 99.92 \\
              
              & MBCNN+CCI ($A_m$) & 99.97 & 99.93 & 99.95 \\
              
              & MBCNN+CCI ($A_r$) & \textbf{99.98} & 99.92 & 99.95 \\
              
              & MBCNN+CCI ($A_m$ \& $A_r$) & 99.95 & \textbf{99.95} & \textbf{99.95} \\
\hline\hline
\end{tabular}
\end{table}

\begin{table}[h]
\centering
\caption{The $F_1$ results of common networks used as Encoders with and without CCI for QRS location. The better results are \textbf{bold-faced}.}
\label{tab:2}
\begin{tabular}{ccccc}
\hline

\renewcommand{\arraystretch}{1.25}
\setlength{\tabcolsep}{1.2mm}

Database      & CPSC2019-Test & MITDB & INCART & QT    \\
\hline\hline
DenseNet      & 98.86 & 99.22 & 99.10 & 99.80\\
DenseNet+CCI  & \textbf{99.11} & \textbf{99.36} & \textbf{99.28} & \textbf{99.93}\\

\cline{1-5}
TCN           & 99.04 & 99.44 & 99.22 & 99.85 \\
TCN+CCI       & \textbf{99.20} & \textbf{99.47} & \textbf{99.37} & \textbf{99.91} \\

\cline{1-5}
SE-TCN         & 99.09 & 99.47 & 99.36 & 99.88 \\
SE-TCN+CCI     & \textbf{99.31} & \textbf{99.51} & \textbf{99.44} & \textbf{99.93} \\
              
\hline\hline
\end{tabular}
\end{table}

\paragraph{Results for QRS location} Table \ref{tab:1} represents the performance of the same model training with CCI on each attribute ($A_r$ or $A_m$) and both attributes.
Here we evaluated the proposed assumption on the four independent and classic databases, CPSC2019-Test, MITDB, INCART and QT.
According to the results of the ablation study, intervention on the morphology attribute ($A_m$) and the rhythm attribute ($A_r$) in the latent space is effective and superimposed, which confirms our assumption on SCM with abstract attributes.
We see steady gains when training model with CCI.
The backbone model has reached a bottleneck in performance on most databases, yet for long-term ECGs, improvement of 0.1\% on $F_1$ may be equivalent to reducing hundreds or thousands of $FP$s and $FN$s.
This can immensely alleviate the workload of cardiologists and reduce the cumulative burden of errors in subsequent diagnostics.
The improvement of performance brought by CCI is mainly reflected in complex ECGs.

We show four typical examples with severe pathological variation and noise contamination in Fig. \ref{fig:4}.
It is apparent to see that the model trained with CCI can significantly reduce errors when recognizing variant QRS-complex or QRSized noise.
Meanwhile, CCI can also weaken the response to the repeated P wave pattern for ECGs with severe auriculo-ventricular block, in which the relative position of the P wave and QRS-complex is unfixed.

Table \ref{tab:2} summarizes the performance gain brought by CCI for different Encoders for QRS location.
From these results, it can be seen that CCI is effective for the common network architectures, which is consistent with the tendency when using MBCNN as the Encoder.
However, we also noticed that the performance gain is not sufficient when the Encoder is not ideally fitted.

\begin{table}[h]
  \centering
  \caption{Ablation results (\%) for heart sound segmentation with MBCNN as the backbone Encoder. The results are average of five sub models under 5-fold cross validation and the better results are \textbf{bold-faced}.}
  \label{tab:3}
  \renewcommand{\arraystretch}{1.25}
  \setlength{\tabcolsep}{1.8mm}
  \begin{tabular}{lllll}
    \hline
    Database & Method & $Se$ & $P_+$ & $F_1$ \\
    \hline\hline
    \multirow{3}{*}{Set-A} & MBCNN & 95.76 & 94.39 & 95.07 \\
    
    & MBCNN+CCI ($A_m$) & \textbf{96.24} & 95.71 & 95.97 \\
    
    & MBCNN+CCI ($A_r$) & 95.92 & 95.59 & 95.76 \\
    
    & MBCNN+CCI ($A_m$ \& $A_r$) & 96.10 & \textbf{95.96} & \textbf{96.03} \\
    
    \cline{1-5}
    
    \multirow{3}{*}{Set-B} & MBCNN & 87.91 & 88.28 & 88.09 \\
    
    & MBCNN+CCI ($A_m$) & \textbf{89.32} & 90.22 & 89.77 \\
    
    & MBCNN+CCI ($A_r$) & 88.58 & 89.99 & 89.28 \\
    
    & MBCNN+CCI ($A_m$ \& $A_r$) & 89.20 & \textbf{90.72} & \textbf{89.96} \\
    
    \cline{1-5}
    
    \multirow{3}{*}{Set-C} & MBCNN & 91.39 & 89.41 & 90.39 \\
    
    & MBCNN+CCI ($A_m$) & 92.10 & 91.03 & 91.56 \\
    
    & MBCNN+CCI ($A_r$) & 91.86 & 91.03 & 91.45 \\
    
    & MBCNN+CCI ($A_m$ \& $A_r$) & \textbf{92.37} & \textbf{91.72} & \textbf{92.04} \\
    
    \cline{1-5}
    
    \multirow{3}{*}{Set-D} & MBCNN & 95.80 & 94.20 & 94.99 \\
    
    & MBCNN+CCI ($A_m$) & \textbf{96.07} & 94.93 & \textbf{95.49} \\
    
    & MBCNN+CCI ($A_r$) & 95.94 & 94.89 & 95.41 \\
    
    & MBCNN+CCI ($A_m$ \& $A_r$) & 95.93 & \textbf{94.99} & 95.45 \\
    
    \cline{1-5}
    
    \multirow{3}{*}{Set-E} & MBCNN & 91.45 & 94.28 & 92.84 \\
    
    & MBCNN+CCI ($A_m$) & \textbf{92.76} & 95.62 & 94.17 \\
    
    & MBCNN+CCI ($A_r$) & 92.57 & 95.61 & 94.07 \\
    
    & MBCNN+CCI ($A_m$ \& $A_r$) & 92.65 & \textbf{96.11} & \textbf{94.34} \\
    
    \cline{1-5}
    
    \multirow{3}{*}{Set-F} & MBCNN & 84.90 & 83.78 & 84.33 \\
    
    & MBCNN+CCI ($A_m$) & 84.13 & 85.74 & 84.93 \\
    
    & MBCNN+CCI ($A_r$) & 85.38 & 87.40 & 86.38 \\
    
    & MBCNN+CCI ($A_m$ \& $A_r$) & \textbf{85.58} & \textbf{88.58} & \textbf{87.06} \\
    
    \cline{1-5}

    \multirow{3}{*}{Set-G} & MBCNN & 89.66 & 88.93 & 89.29  \\
    
    & MBCNN+CCI ($A_m$) & \textbf{90.11} & 90.71 & 90.41 \\
    
    & MBCNN+CCI ($A_r$) & 89.51 & 90.26 & 89.88 \\
    
    & MBCNN+CCI ($A_m$ \& $A_r$) & 89.92 & \textbf{91.30} & \textbf{90.60} \\
    
    \cline{1-5}
    
    \multirow{3}{*}{Set-H} & MBCNN & 92.29 & 90.14 & 92.17 \\
    
    & MBCNN+CCI ($A_m$) & 94.15 & 91.54 & 92.82 \\
    
    & MBCNN+CCI ($A_r$) & 93.41 & 91.66 & 92.53 \\
    
    & MBCNN+CCI ($A_m$ \& $A_r$) & \textbf{94.36} & \textbf{92.36} & \textbf{93.35} \\
    
    \hline\hline
  \end{tabular}
\end{table}

\paragraph{Results of heart sound segmentation} We reports the evaluation metrics of the model training with and without CCI on databases from PhysioNet/CinC Challenge 2016 in Table \ref{tab:3}.
Similar observations to QRS location can be obtained.
For heart sound segmentation, the improvement of $F_1$ induced by CCI is more significant.
On all the sub databases, the model training with CCI outperforms the backbone method by at least 1.0\% on most databases, even 2.73\% on Set-F.
Moreover, CCI causes a more consistent $F_1$ performance when segmenting different states.
Also we can conclude that on the whole sub databases, the performance is further improved when we implement CCI with both attributes, $A_r$ and $A_m$ expect for Set-D.
Such as on Set-F, intervention on both attributes improves the $F_1$ by 0.5\%\textasciitilde 2\% compared to intervention on $A_r$ and $A_m$ solely.

\subsubsection{Q2: Test with SNR controllered samples}
In this section, we mainly analyze the changes brought by CCI as a constraint while training for the robustness of segmenting cardiovascular signals.
A noise stress test with typical noises of ambulatory ECG records for QRS location and lung sounds for heart sound segmentation is conducted, also with different degrees of intra-band Gaussian noise for both tasks.
We tested all the sub-models from 5-fold evaluation under different noise types and signal-to-noise ratios (SNR), and calculated the mean and standard deviation of the corresponding error rates.
Test examples were generated from database CPSC2019-Test for QRS location and Training-a (other than the records for training) for heart sound segmentation by adding different types of noise globally with controlled SNR.

\begin{figure}[t]
    \begin{center}
    \subfigure[Lung sound]{
    \includegraphics[width=4cm]{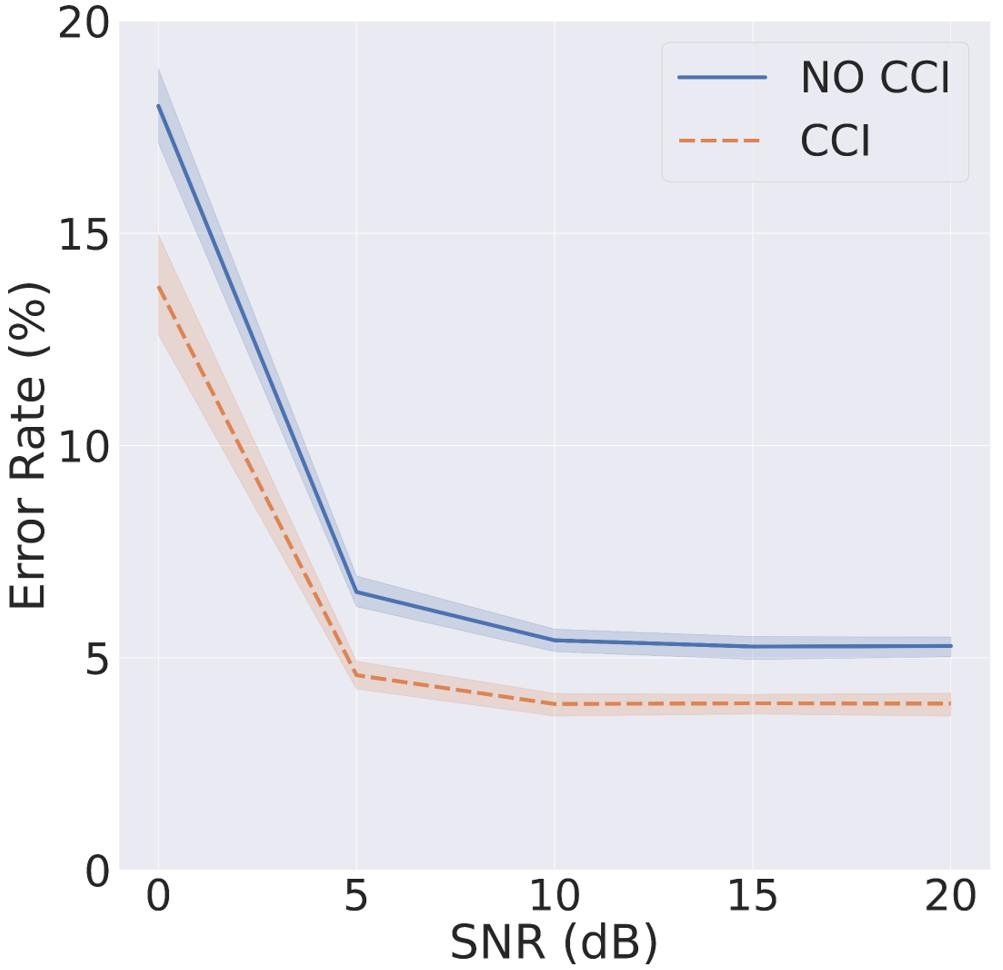}
    }
    \subfigure[Intra-band Gaussian noise]{
    \includegraphics[width=4cm]{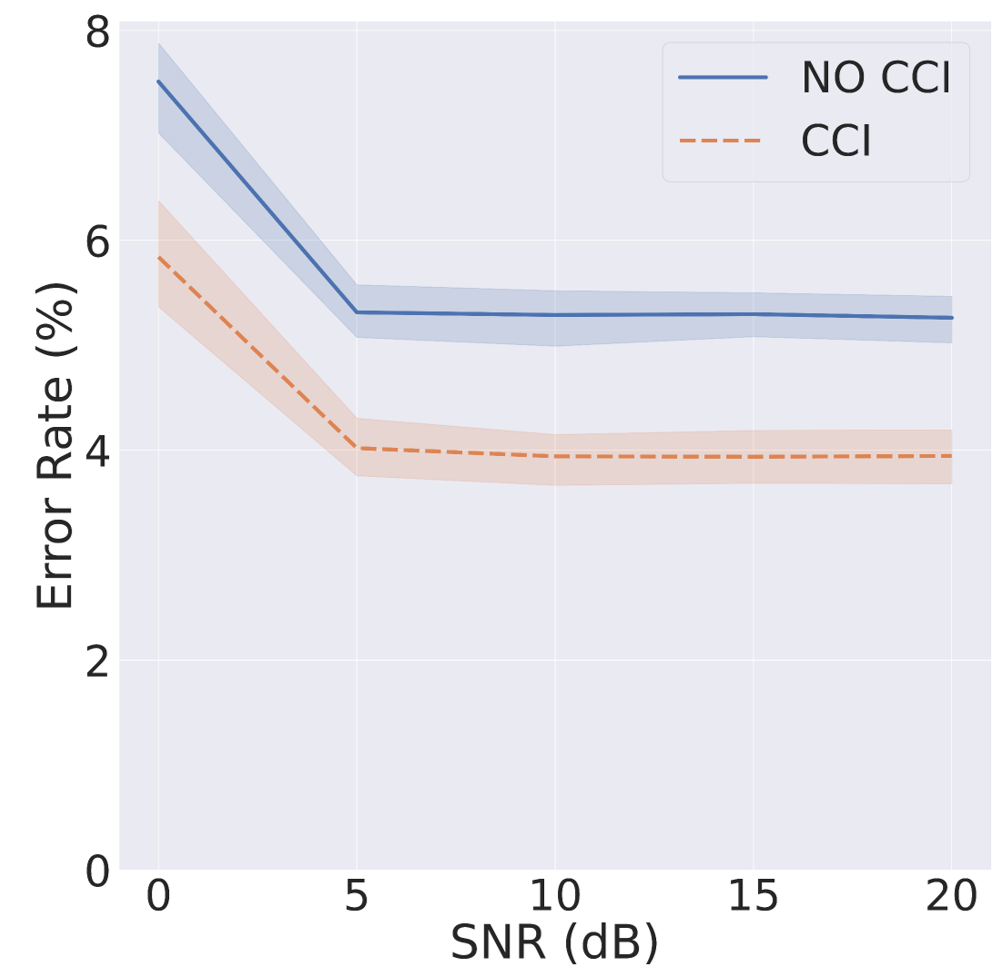}
    }
    \caption{The results of noise stress test (nst) for CCI in heart sound segmentation task.}
    \label{fig:5}
    \end{center}
\end{figure}

\paragraph{Main Results}
Fig. \ref{fig:4} and Fig. \ref{fig:5} shows the error rates at different SNR for the segmentation model training with and without CCI.
It is evident that the model training with CCI has the highest performance in all noise levels.
CCI also results in the slowest performance decay compared to the backbone method.
Among all the noise categories, muscle artifact (ma) and intra-band Gaussian noise have the greatest influence on QRS location performance.
Especially intra-band Gaussian noise causes nearly 50\% increase in $Er$ at 0dB SNR.
Regardless of noise type, the model training with CCI maintains the lowest $Er$ and the most stable performance in the comparison of backbone method.
Expect for intra-band Gaussian noise, CCI steadily reduces the $Er$ by 2\%-5\% while noise stress testing.
Since intra-band Gaussian noise is more likely to induce morphological interference in ECGs and heart sounds, CCI only reduces the $Er$ by about 1\%.
This also illustrates the importance of $A_m$ in the segmentation of cardiovascular signals.

\begin{figure}[t]
    \begin{center}
    \subfigure[Baseline wander]{
    \includegraphics[width=4cm]{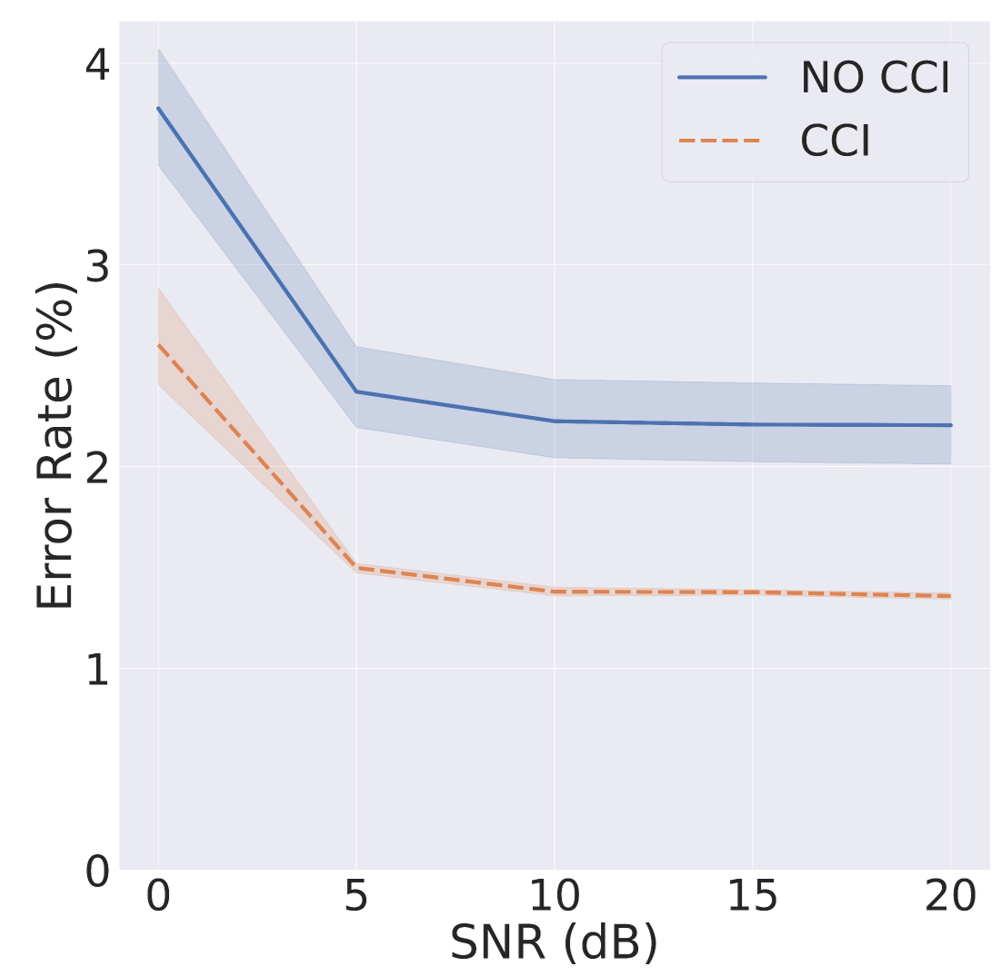}
    }
    \subfigure[Electronic motion]{
    \includegraphics[width=4cm]{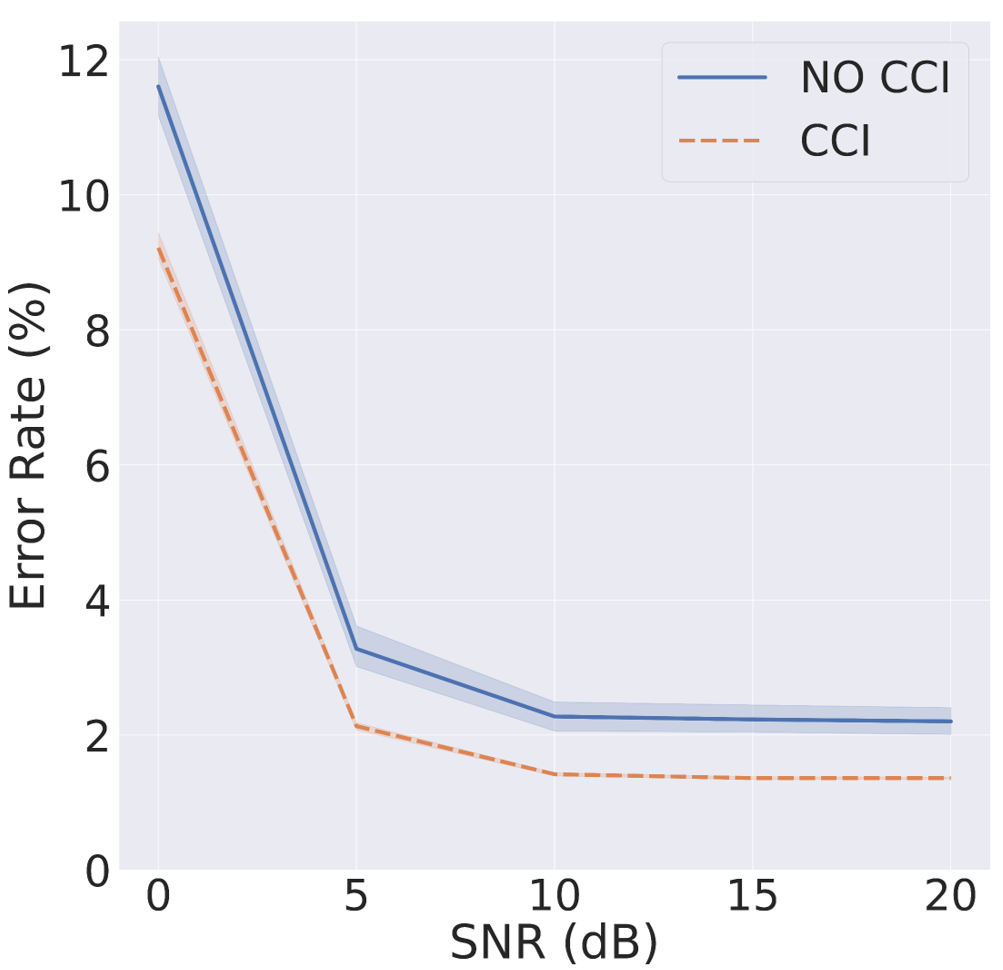}
    } \\
    \subfigure[Muscle artifact]{
    \includegraphics[width=4cm]{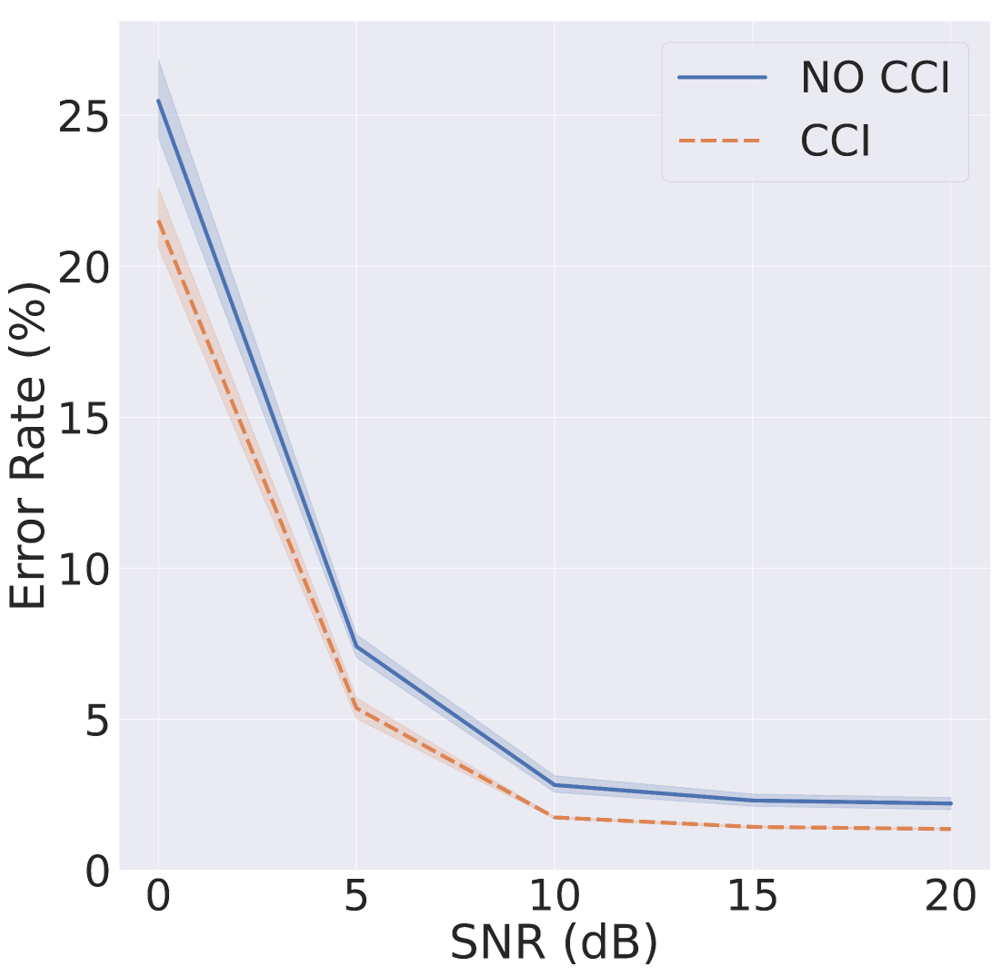}
    }
    \subfigure[Intra-band Gaussian noise]{
    \includegraphics[width=4cm]{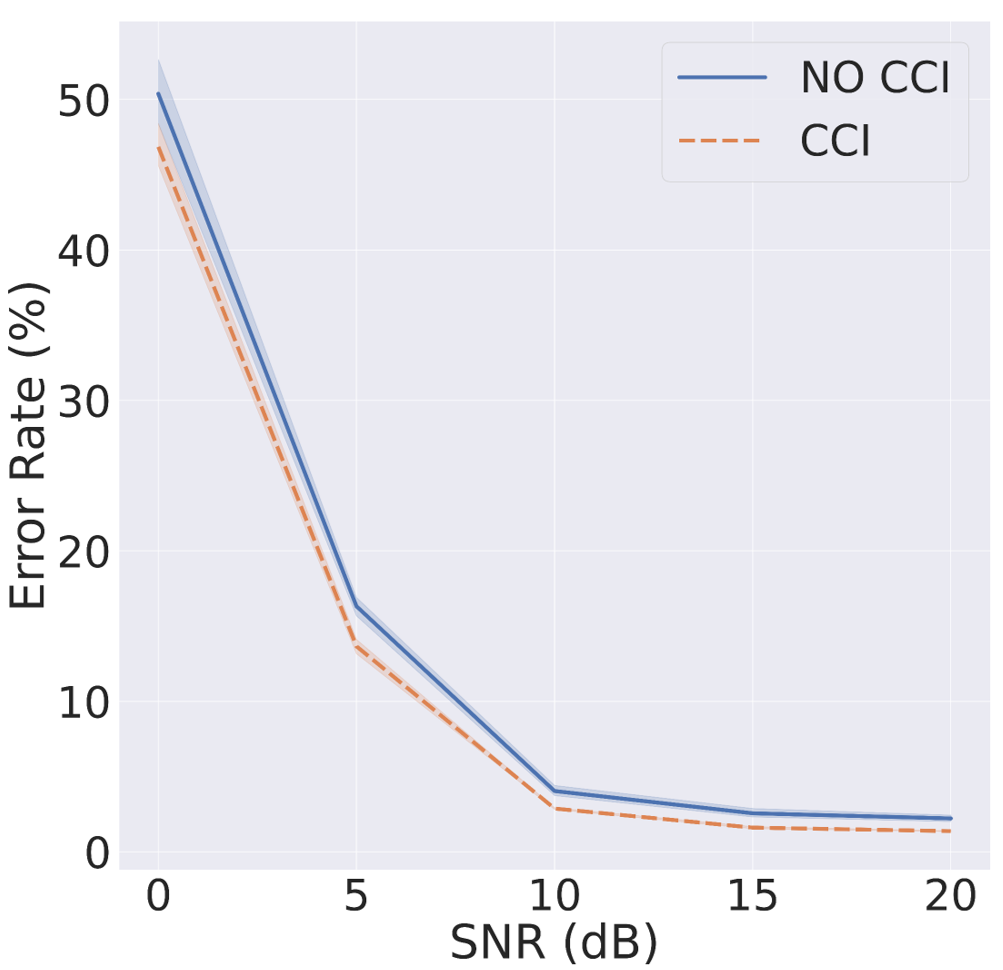}
    }
\caption{The results of noise stress test (nst) for CCI in QRS location task.}
    \label{fig:4}
    \end{center}
\end{figure}

\begin{figure*}[ht]
    \begin{center}
    \subfigure[Representations generated by MBCNN trained without CCI.]{
    \label{fig:6a}
    \includegraphics[width=4.5cm]{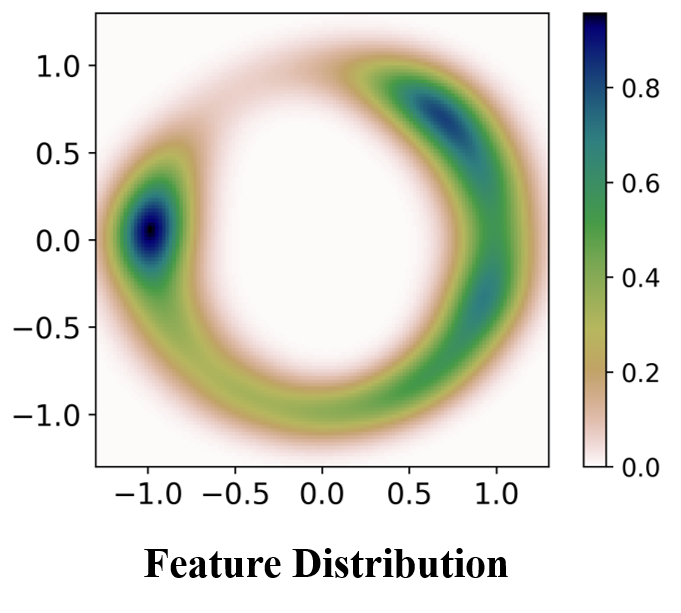}
    \includegraphics[width=4.5cm]{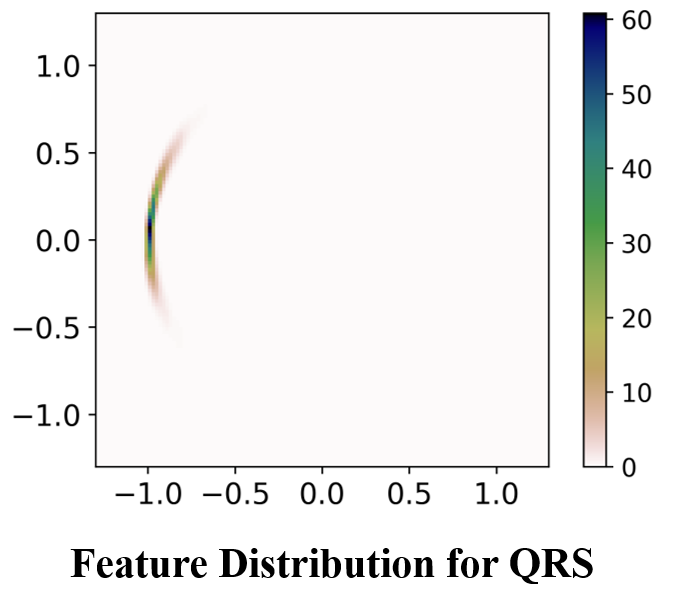}
    \includegraphics[width=4.5cm]{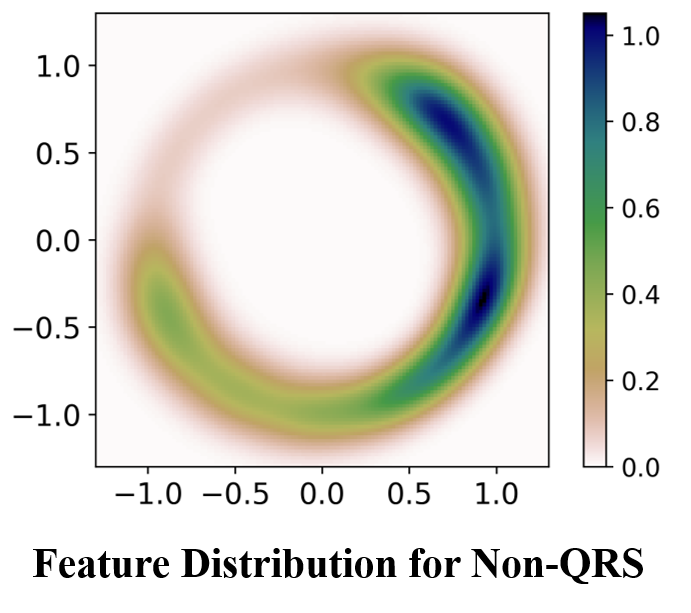}
    } \\
    \subfigure[Representations generated by MBCNN trained with CCI.]{
    \label{fig:6b}
    \includegraphics[width=4.5cm]{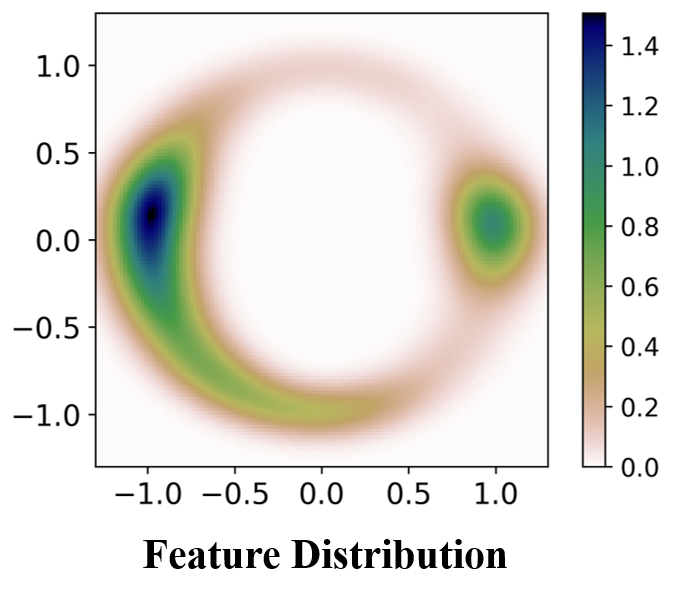}
    \includegraphics[width=4.5cm]{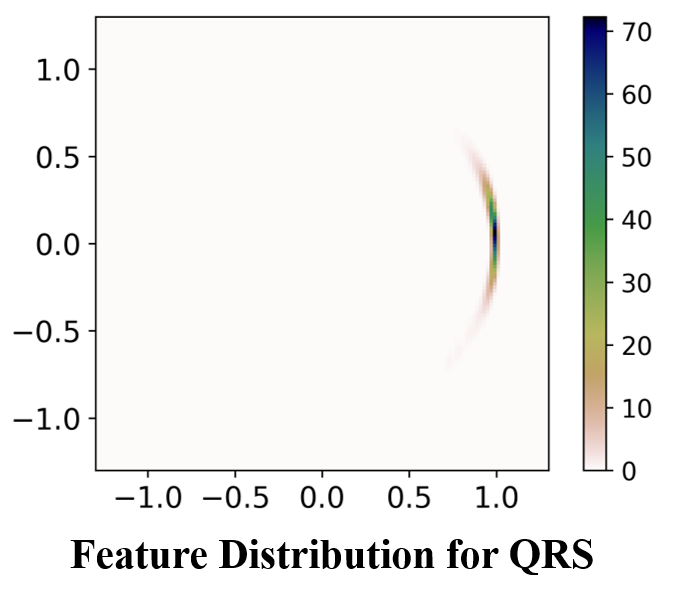}
    \includegraphics[width=4.5cm]{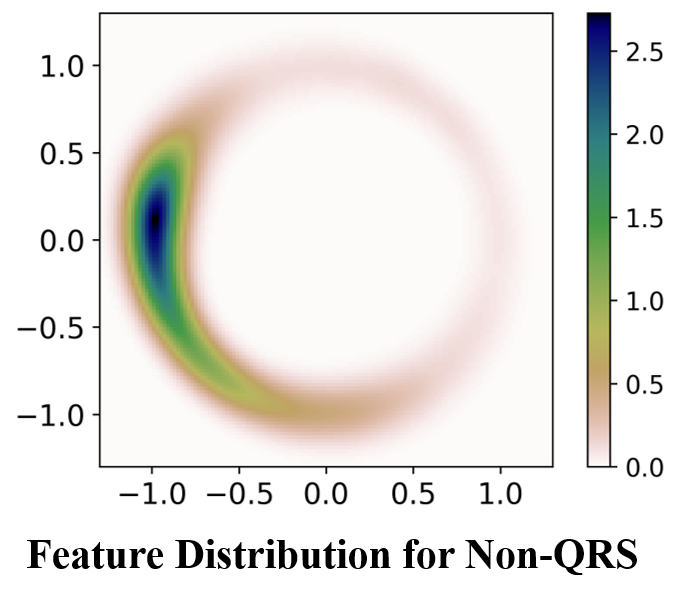}
    } \\
    \caption{Representation diversity of training set for QRS-complex location in $\mathbb{R}^2$ with Gaussian kernel density estimation (KDE).}
    \label{fig:6}
    \end{center}
\end{figure*}

\begin{figure*}[!t]
    \begin{center}
    \subfigure[Representations generated by MBCNN trained without CCI.]{
    \label{fig:7a}
    \includegraphics[width=3.5cm]{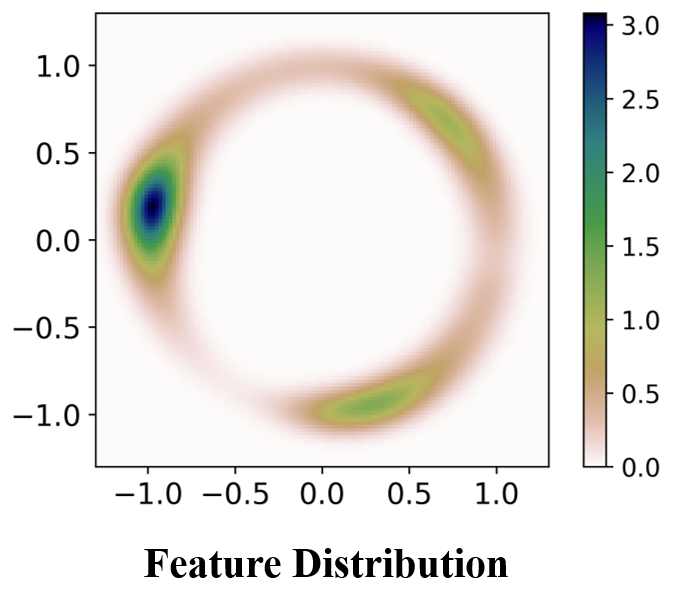}
    \includegraphics[width=3.5cm]{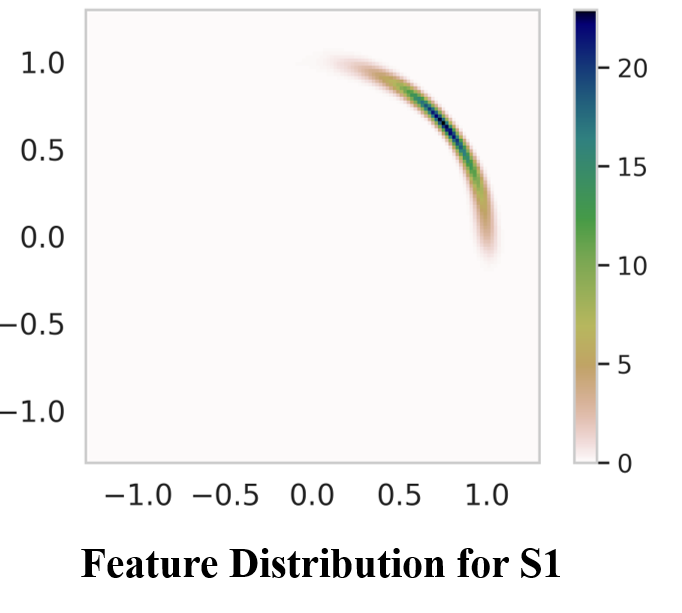}
    \includegraphics[width=3.5cm]{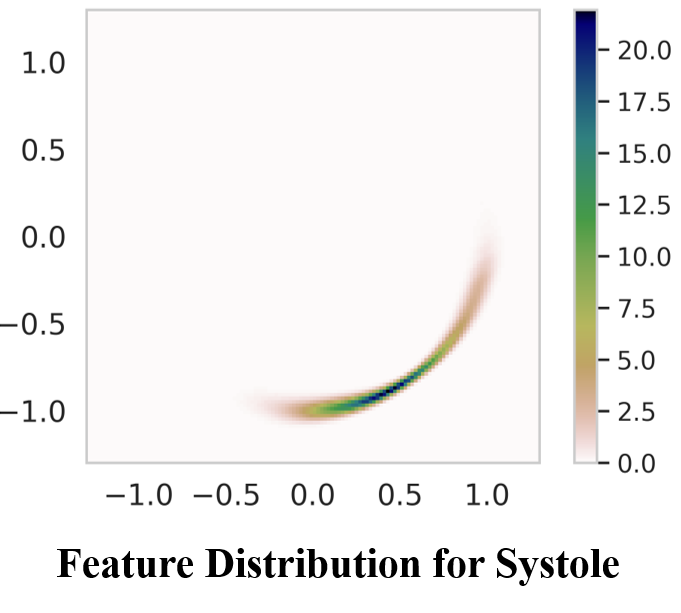}
    \includegraphics[width=3.5cm]{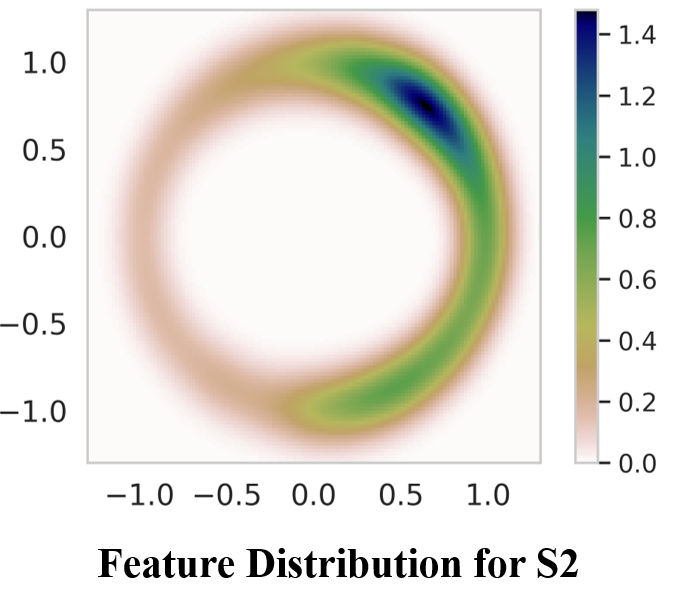}
    \includegraphics[width=3.5cm]{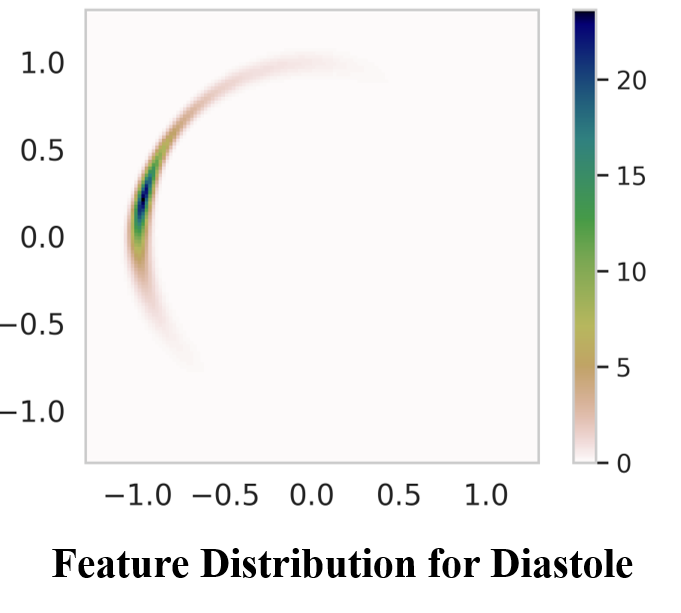}
    } \\
    \subfigure[Representations generated by MBCNN trained with CCI.]{
    \label{fig:7b}
    \includegraphics[width=3.5cm]{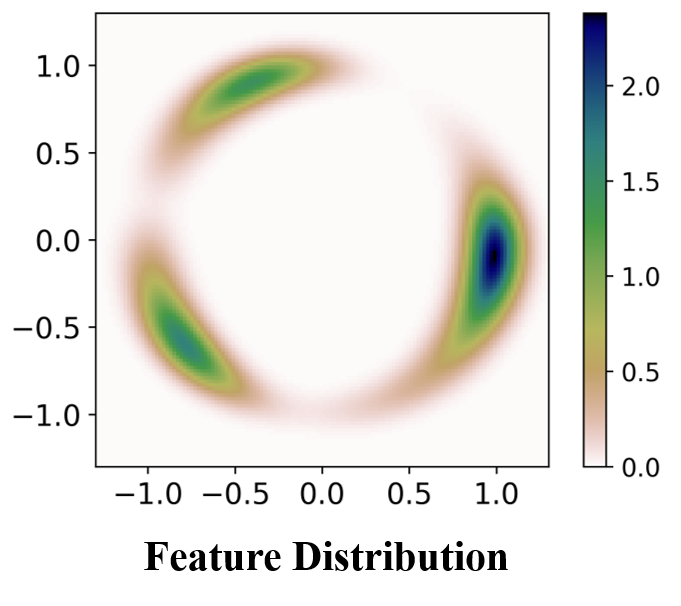}
    \includegraphics[width=3.5cm]{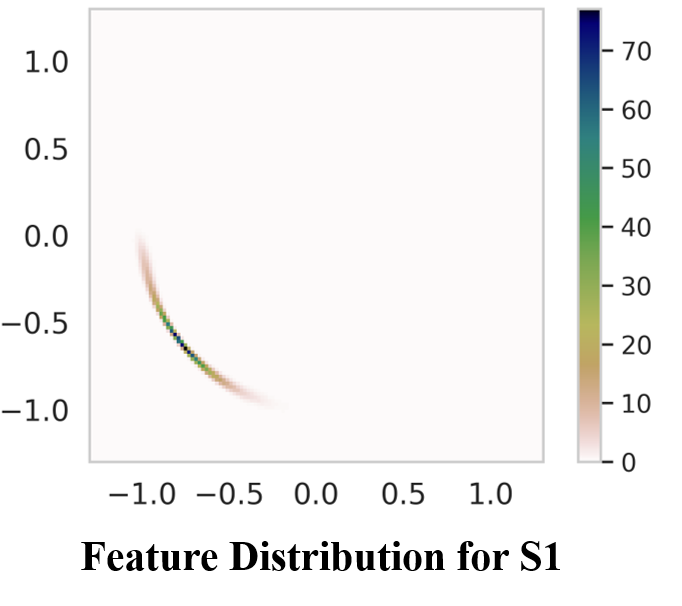}
    \includegraphics[width=3.5cm]{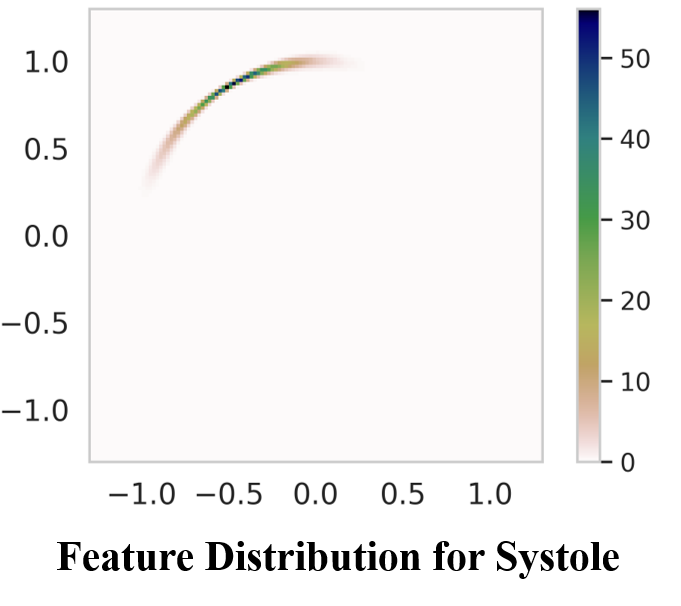}
    \includegraphics[width=3.5cm]{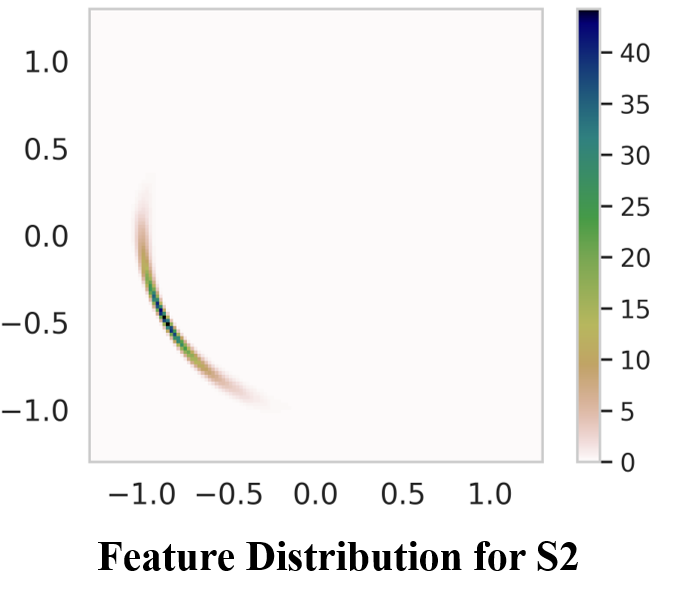}
    \includegraphics[width=3.5cm]{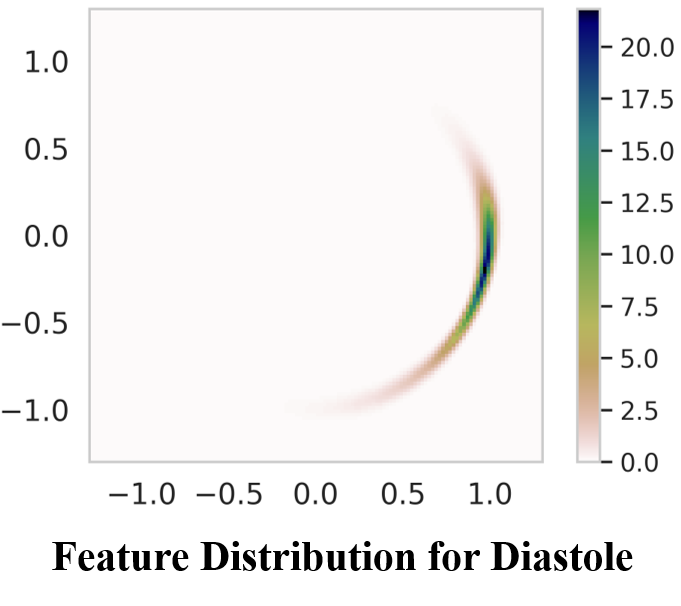}
    } 
    \caption{Representation diversity of training set for heart sound segmentation in $\mathbb{R}^2$ with Gaussian kernel density estimation (KDE).}
    \label{fig:7}
    \end{center}
\end{figure*}

\subsection{Q3: Visualization of Feature Density}

According to the previous assumption, CCI should eliminate the implicit statistical bias brought by the single attribute and lead to more objective representations.
For the conventional training, $A_m$ and $A_r$ may confounding the Encoder in distilling the intrinsic factors for state discrimination.
Therefore, the feature representation learned with CCI ought to be more concentrated.
To empirically verify this, we compress the deep features encoded by MBCNN to the unit hypersphere to visualize the latent distribution of different states.
The representations of each state are grouped by frames with interval of 16ms for QRS location and 100ms for heart sound segmentation.
A Gaussian kernel with bandwidth estimated by Scott's Rule \cite{scott2015multivariate} is applied to estimate the probability density function of the generated representations after dimensionality reduction with principle component analysis (PCA) and normalization.
Darker areas have more concentrated features, and if the feature space (the 2-dim sphere) is covered by dark areas, it has more diversely placed features. 

The visualization results are shown in Fig. \ref{fig:6} and Fig. \ref{fig:7}.
It can be observed that, with CCI, the deep features of different states form more tight and concentrated clusters. 
Intuitively, they are potentially more separable from each other.
In contrast, features learned without CCI are distributed in clusters that have more overlapped parts.
The evident discrepancy occurs in non-QRS representations for QRS location and S2 representations for heart sound segmentation.
This demonstrate why CCI improves the segmentation performance to a certain extent.

\section{Discussion and Conclusion}
\label{sec:discussion and conclusion}
In this work, we introduce a contrastive causal intervention scheme (CCI) for learning semantic representations of cardiovascular signals.
CCI is a frame-level constraint for the training process to eliminate the implicit confounding factors induced by $A_m$ and $A_r$.
We show that training with CCI can effectively improve the segmentation performance and adapt to other independent databases and various networks.
Furthermore, the proposed method is considerably efficient to train a segmentation model generalizing to noisy cardiovascular signals.
According to the visualization results of the latent distribution encoded with and without CCI, it is sensible to attribute the performance gain to a more separable and state-concentrated deep feature space brought by CCI.

Contrastive learning has been out-standingly successful for CV and NLP, especially in self-supervised tasks.
However, the existing work of contrastive learning in CV and NLP seems inappropriate to be applied for cardiovascular signals.
For example, in CV and NLP, masking a part of the data and aligning the representation of the masked area with the original representation is a common framework.
Yet for event-based analysis of cardiovascular signals, the disappearance of heartbeats may correspond to cardiac arrest, not a noise masking.
In this work, we propose a contrastive learning framework based on the causal attributes of cardiovascular signals, and summarize several suggestions for further exploring.

1) \textbf{If it is effective to construct a causal graph of the inner attributes of the data, how can we explore the intrinsic causality of more complex task with cardiovascular signals?}
Our assumption of causal intervention on rhythm and morphology attribution was based on a prior intuition, corresponding to the cognition when we identify each state in a cardiac cycle.
However, except for semantic segmentation, the classification of cardiovascular signals requires the abstracted causal mechanism more detailed.
One possible direction is to establish the preliminary research on the binary classification task, like diagnosis of atrial fibrillation.
Other attribute such as Markov chain of state transition is also a critical causal dependency when we doing deep representation learning for cardiovascular signals.

2) \textbf{Excluding the interference of confounders in a specific task, should a better concentrated representation be obtained?}
In \cite{wang2020understanding}, the researchers have presented a connection between contrastive loss and the alignment and uniformity properties.
The analysis is set in image classification and unsupervised learning.
In the visualization results of this work, we have found out that when utilizing CCI in training process, the generated representations of different states are more aligned, less uniformed.
As the organizer of CPSC2019 and the trimmer of PhysioNet/CinC Challenge 2016, we understand that the annotations of the training data from CPSC2019 and Training-a can be 100\% confident through contextual information.
Therefore, it is reasonable to have more concentrated representations when we reduce the confounding impact.
Nonetheless, the cardiovascular signals own low frequency band and are always highly uncertain while testing due to variation and noise contamination.
The research of whether the introduction of representation uniformity can measure and distinguish the uncertain state is a worthy investment.

\bibliographystyle{IEEEtran}
\bibliography{references}

\vfill

\end{document}